\documentstyle{article}
\makeatletter
\def\lop#1\to#2{\expandafter\lopoff#1\lopoff#1#2}
\long\def\lopoff,#1,#2\lopoff#3#4{\def#4{#1}\def#3{,#2}}
\def\@@mlistempty{,}
\newif\iflistnonempty
\def\multiputlist(#1,#2)(#3,#4){\@ifnextchar
[{\@imultiputlist(#1,#2)(#3,#4)}{\@imultiputlist(#1,#2)(#3,#4)[]}}

\long\def\@imultiputlist(#1,#2)(#3,#4)[#5]#6{{%
\@xdim=#1\unitlength \@ydim=#2\unitlength
\listnonemptytrue \def\@@mlist{,#6,} % need this for end condition
\loop
\lop\@@mlist\to\@@firstoflist
\@killglue\raise\@ydim\hbox to\z@{\hskip
\@xdim\@imakepicbox(0,0)[#5]{\@@firstoflist}\hss}
\advance\@xdim #3\unitlength\advance\@ydim #4\unitlength
\ifx\@@mlist\@@mlistempty \listnonemptyfalse\fi
\iflistnonempty
\repeat\relax
\ignorespaces}}
\newcount\@@multicnt
\def\matrixput(#1,#2)(#3,#4)#5(#6,#7)#8#9{%
\ifnum#5>#8\@matrixput(#1,#2)(#3,#4){#5}(#6,#7){#8}{#9}%
\else\@matrixput(#1,#2)(#6,#7){#8}(#3,#4){#5}{#9}\fi}

\long\def\@matrixput(#1,#2)(#3,#4)#5(#6,#7)#8#9{{\@killglue%
\@multicnt=#5\relax\@@multicnt=#8\relax%
\@xdim=0pt%
\@ydim=0pt%
\setbox\@tempboxa\hbox{\@whilenum \@multicnt > 0\do {%
\raise\@ydim\hbox to \z@{\hskip\@xdim #9\hss}%
\advance\@multicnt \m@ne%
\advance\@xdim #3\unitlength\advance\@ydim #4\unitlength}}%
\@xdim=#1\unitlength%
\@ydim=#2\unitlength%
\@whilenum \@@multicnt > 0\do {%
\raise\@ydim\hbox to \z@{\hskip\@xdim \copy\@tempboxa\hss}%
\advance\@@multicnt \m@ne%
\advance\@xdim #6\unitlength\advance\@ydim #7\unitlength}%
\ignorespaces}}
\newcount\d@lta
\newdimen\@delta
\newdimen\@@delta
\newcount\@gridcnt
\def\grid(#1,#2)(#3,#4){\@ifnextchar [{\@igrid(#1,#2)(#3,#4)}%
{\@igrid(#1,#2)(#3,#4)[@,@]}}

\long\def\@igrid(#1,#2)(#3,#4)[#5,#6]{%
\makebox(#1,#2){%
\@delta=#1pt\@@delta=#3pt\divide\@delta \@@delta\d@lta=\@delta%
\advance\d@lta \@ne\relax\message{grid=\the\d@lta\space x}%
\multiput(0,0)(#3,0){\d@lta}{\hbox to\z@{\hskip -\@halfwidth \vrule
	 \@width \@wholewidth \@height #2\unitlength \@depth \z@\hss}}%
\ifx#5@\relax\else%
\global\@gridcnt=#5%
\multiput(0,0)(#3,0){\d@lta}{%
\makebox(0,-2)[t]{\number\@gridcnt\global\advance\@gridcnt by #3}}%
\global\@gridcnt=#5%
\multiput(0,#2)(#3,0){\d@lta}{\makebox(0,0)[b]{\number\@gridcnt\vspace{2mm}%
\global\advance\@gridcnt by #3}}%
\fi%
\@delta=#2pt\@@delta=#4pt\divide\@delta \@@delta\d@lta=\@delta%
\advance\d@lta \@ne\relax\message{\the\d@lta . }%
\multiput(0,0)(0,#4){\d@lta}{\vrule \@height \@halfwidth \@depth \@halfwidth
	 \@width #1\unitlength}%
\ifx#6@\relax\else
\global\@gridcnt=#6%
\multiput(0,0)(0,#4){\d@lta}{%
\makebox(0,0)[r]{\number\@gridcnt\ \global\advance\@gridcnt by #4}}%
\global\@gridcnt=#6%
\multiput(#1,0)(0,#4){\d@lta}{%
\makebox(0,0)[l]{\ \number\@gridcnt\global\advance\@gridcnt by #4}}%
\fi}}
\def\picsquare{\hskip -0.5\@wholewidth%
\vrule height \@halfwidth depth \@halfwidth width \@wholewidth}
\def\picsquare@bl{\vrule height \@wholewidth depth \z@  width \@wholewidth}
\newif\if@jointhem \global\@jointhemfalse
\newif\if@firstpoint \global\@firstpointtrue
\newcount\@joinkind
\def\dottedjoin{\global\@jointhemtrue \global\@joinkind=0\relax
  \bgroup\@ifnextchar[{\@idottedjoin}{\@idottedjoin[\picsquare@bl]}}
\def\@idottedjoin[#1]#2{\gdef\dotchar@join{#1}\gdef\dotgap@join{#2}}
\def\enddottedjoin{\global\@jointhemfalse \global\@firstpointtrue\egroup}
\def\dashjoin{\global\@jointhemtrue \global\@joinkind=1\relax
  \bgroup\@ifnextchar[{\@idashjoin}{\@idashjoin[\dashlinestretch]}}
\def\@idashjoin[#1]#2{\edef\dashlinestretch{#1}\gdef\dashlen@join{#2}%
\@ifnextchar[{\@iidashjoin}{\gdef\dotgap@join{}}}
\def\@iidashjoin[#1]{\gdef\dotgap@join{#1}}

\def\drawjoin{\global\@jointhemtrue \global\@joinkind=2\relax
  \bgroup\@ifnextchar[{\@idrawjoin}{}}
\def\@idrawjoin[#1]{\def\drawlinestretch{#1}}

\long\def\jput(#1,#2)#3{{\@killglue\raise#2\unitlength\hbox to \z@{\hskip
#1\unitlength #3\hss}\ignorespaces}
\if@jointhem
 \if@firstpoint \gdef\x@one{#1} \gdef\y@one{#2} \global\@firstpointfalse
 \else\ifcase\@joinkind
	\@dottedline[\dotchar@join]{\dotgap@join\unitlength}%
(\x@one\unitlength,\y@one\unitlength)(#1\unitlength,#2\unitlength)
	\or\@dashline[\dashlinestretch]{\dashlen@join}[\dotgap@join]%
(\x@one,\y@one)(#1,#2)
	\else\@drawline[\drawlinestretch](\x@one,\y@one)(#1,#2)\fi
    \gdef\x@one{#1} \gdef\y@one{#2}
 \fi
\fi}
\newdimen\@dotgap
\newdimen\@ddotgap
\newcount\@x@diff
\newcount\@y@diff
\newdimen\x@diff
\newdimen\y@diff
\newbox\@dotbox
\newcount\num@segments
\newcount\num@segmentsi
\newif\ifsqrt@done
\def\sqrtandstuff#1#2#3{
\ifdim #1 <0pt \@x@diff= -#1 \else\@x@diff=#1\fi
\ifdim #2 <0pt \@y@diff= -#2 \else\@y@diff=#2\fi
\@dotgap=#3 \divide\@dotgap \tw@
\advance\@x@diff \@dotgap \advance\@y@diff \@dotgap% for round-off errors
\@dotgap=#3
\divide\@x@diff \@dotgap \divide\@y@diff \@dotgap
\sqrt@donefalse
\ifnum\@x@diff < 2
   \ifnum\@y@diff < 2 \num@segments=\@x@diff \advance\num@segments \@y@diff
		      \sqrt@donetrue
        \else\num@segments=\@y@diff \sqrt@donetrue\fi
   \else\ifnum\@y@diff < 2 \num@segments=\@x@diff \sqrt@donetrue\fi
\fi
\ifsqrt@done \ifnum\num@segments=\z@ \num@segments=\@ne\fi\relax
 \else \ifnum\@y@diff >\@x@diff
		 \@tempcnta=\@x@diff \@x@diff=\@y@diff \@y@diff=\@tempcnta
       \fi    		%exchange @x@diff & @y@diff, so now @x@diff > @y@diff
  \num@segments=\@y@diff
  \multiply\num@segments \num@segments
  \multiply\num@segments by 457
  \divide\num@segments \@x@diff
  \advance\num@segments by 750 % for round-off, going to divide by 1000.
  \divide\num@segments \@m
  \advance\num@segments \@x@diff
\fi}
\def\dottedline{\@ifnextchar [{\@idottedline}{\@idottedline[\picsquare@bl]}}
\def\@idottedline[#1]#2(#3,#4){\@ifnextchar (%
{\@iidottedline[#1]{#2}(#3,#4)}{\relax}}
\def\@iidottedline[#1]#2(#3,#4)(#5,#6){\@dottedline[#1]{#2\unitlength}%
(#3\unitlength,#4\unitlength)(#5\unitlength,#6\unitlength)%
\@idottedline[#1]{#2}(#5,#6)}
\long\def\@dottedline[#1]#2(#3,#4)(#5,#6){{%
\x@diff=#5\relax\advance\x@diff by -#3\relax
\y@diff=#6\relax\advance\y@diff by -#4\relax
\sqrtandstuff{\x@diff}{\y@diff}{#2}
\divide\x@diff \num@segments
\divide\y@diff \num@segments
\advance\num@segments \@ne     % to put the last point at destination.
\setbox\@dotbox\hbox{#1}% just to get the dimensions of the character.
\@xdim=#3 \@ydim=#4
\ifdim\ht\@dotbox >\z@% otherwise its a circle.
  \advance\@xdim -0.5\wd\@dotbox
  \advance\@ydim -0.5\ht\@dotbox
  \advance\@ydim .5\dp\@dotbox\fi
\@killglue
\loop \ifnum\num@segments > 0
\unskip\raise\@ydim\hbox to\z@{\hskip\@xdim #1\hss}%
\advance\num@segments \m@ne\advance\@xdim\x@diff\advance\@ydim\y@diff%
\repeat
\ignorespaces}}
\def\dashlinestretch{0} %well, could have used a counter.
\def\dashline{\@ifnextchar [{\@idashline}{\@idashline[\dashlinestretch]}}
\def\@idashline[#1]#2{\@ifnextchar [{\@iidashline[#1]{#2}}%
{\@iidashline[#1]{#2}[\@empty]}} %\@empty needed-- later checked with \ifx
\def\@iidashline[#1]#2[#3](#4,#5){\@ifnextchar (%
{\@iiidashline[#1]{#2}[#3](#4,#5)}{\relax}}
\def\@iiidashline[#1]#2[#3](#4,#5)(#6,#7){%
\@dashline[#1]{#2}[#3](#4,#5)(#6,#7)%
\@iidashline[#1]{#2}[#3](#6,#7)}
\long\def\@dashline[#1]#2[#3](#4,#5)(#6,#7){{%
\x@diff=#6\unitlength \advance\x@diff by -#4\unitlength
\y@diff=#7\unitlength \advance\y@diff by -#5\unitlength
\@tempdima=#2\unitlength \advance\@tempdima -\@wholewidth
\sqrtandstuff{\x@diff}{\y@diff}{\@tempdima}
\ifnum\num@segments <3 \num@segments=3\fi% min number of dashes I can plot
\@tempdima=\x@diff \@tempdimb=\y@diff
\divide\@tempdimb by\num@segments
\divide\@tempdima by\num@segments
{\ifx#3\@empty \relax
    \ifdim\@tempdima < 0pt \x@diff=-\@tempdima\else\x@diff=\@tempdima\fi
    \ifdim\@tempdimb < 0pt \y@diff=-\@tempdimb\else\y@diff=\@tempdimb\fi
    \ifdim\x@diff < 0.3pt %it's a vertical dashline
           \ifdim\@tempdimb > 0pt
	        \global\setbox\@dotbox\hbox{\hskip -\@halfwidth \vrule
		 \@width \@wholewidth \@height \@tempdimb}
	   \else\global\setbox\@dotbox\hbox{\hskip -\@halfwidth \vrule
		 \@width \@wholewidth \@height\z@ \@depth -\@tempdimb}\fi
       \else\ifdim\y@diff < 0.3pt %it's a horizontal dashline
               \ifdim\@tempdima >0pt
		  \global\setbox\@dotbox\hbox{\vrule \@height \@halfwidth
		 		\@depth \@halfwidth \@width \@tempdima}
		\else\global\setbox\@dotbox\hbox{\hskip \@tempdima
			 \vrule \@height \@halfwidth \@depth \@halfwidth
				 \@width -\@tempdima \hskip \@tempdima}\fi
	    \else\global\setbox\@dotbox\hbox{%
\@dottedline[\picsquare]{0.98\@wholewidth}(0pt,0pt)(\@tempdima,\@tempdimb)}
\fi\fi
\else\global\setbox\@dotbox\hbox{%
\@dottedline[\picsquare]{#3\unitlength}(0pt,0pt)(\@tempdima,\@tempdimb)}
\fi}
\advance\x@diff by -\@tempdima % both have same sign
\advance\y@diff by -\@tempdimb
\@tempdima=\num@segments\@wholewidth \@tempdima=2\@tempdima
\@tempcnta=\@tempdima \@tempdima=#2\unitlength \@tempdimb=0.5\@tempdima
\@tempcntb=\@tempdimb \advance\@tempcnta by \@tempcntb % round-off error
\divide\@tempcnta by\@tempdima \advance\num@segments by -\@tempcnta
\ifnum #1=0 \relax\else\ifnum #1 < -100
  \typeout{***dashline: reduction > -100 percent implies blankness!***}
\else\num@segmentsi=#1 \advance\num@segmentsi by 100
     \multiply\num@segments by\num@segmentsi \divide\num@segments by 100
\fi\fi
\divide\num@segments by 2 % earlier num@segments included 'empty dashes' too.
\ifnum\num@segments >0 % if =0 then don't divide => \x@diff & \y@diff
 \divide\x@diff by\num@segments%   remain same.
 \divide\y@diff by\num@segments
 \advance\num@segments by\@ne %for the last segment for which I subtracted
 \else\num@segments=2 % one at each end.
\fi
\@xdim=#4\unitlength \@ydim=#5\unitlength
\@killglue
\loop \ifnum\num@segments > 0
\unskip\raise\@ydim\hbox to\z@{\hskip\@xdim \copy\@dotbox\hss}%
\advance\num@segments \m@ne\advance\@xdim\x@diff\advance\@ydim\y@diff%
\repeat
\ignorespaces}}
\newif\if@flippedargs
\def\lineslope(#1,#2){%
\ifdim #1 <0pt \@xdim= -#1 \else\@xdim=#1\fi
\ifdim #2 <0pt \@ydim= -#2 \else\@ydim=#2\fi
\ifdim\@xdim >\@ydim \@tempdima=\@xdim \@xdim=\@ydim \@ydim=\@tempdima
\@flippedargstrue\else\@flippedargsfalse\fi% x < y
\ifdim\@ydim >1pt \@tempcnta=\@ydim
            \divide\@tempcnta by 65536% now \@tempcnta=integral part of #1.
            \divide\@xdim \@tempcnta\fi
\ifdim\@xdim <.083333pt \@xarg=1 \@yarg=0
 \else\ifdim\@xdim <.183333pt	\@xarg=6 \@yarg=1
 \else\ifdim\@xdim <.225pt 	\@xarg=5 \@yarg=1
 \else\ifdim\@xdim <.291666pt 	\@xarg=4 \@yarg=1
 \else\ifdim\@xdim <.366666pt 	\@xarg=3 \@yarg=1
 \else\ifdim\@xdim <.45pt 	\@xarg=5 \@yarg=2
 \else\ifdim\@xdim <.55pt 	\@xarg=2 \@yarg=1
 \else\ifdim\@xdim <.633333pt 	\@xarg=5 \@yarg=3
 \else\ifdim\@xdim <.708333pt 	\@xarg=3 \@yarg=2
 \else\ifdim\@xdim <.775pt 	\@xarg=4 \@yarg=3
 \else\ifdim\@xdim <.816666pt 	\@xarg=5 \@yarg=4
 \else\ifdim\@xdim <.916666pt 	\@xarg=6 \@yarg=5
       \else			\@xarg=1 \@yarg=1%
\fi\fi\fi\fi\fi\fi\fi\fi\fi\fi\fi\fi
\if@flippedargs\relax\else\@tempcnta=\@xarg \@xarg=\@yarg
			  \@yarg=\@tempcnta\fi
\ifdim #1 <0pt \@xarg= -\@xarg\fi
\ifdim #2 <0pt \@yarg= -\@yarg\fi
}
\newif\if@toosmall
\newif\if@drawit
\newif\if@horvline
\def\drawlinestretch{0} %well, could have used a counter.
\def\drawline{\@ifnextchar [{\@idrawline}{\@idrawline[\drawlinestretch]}}
\def\@idrawline[#1](#2,#3){\@ifnextchar ({\@iidrawline[#1](#2,#3)}{\relax}}
\def\@iidrawline[#1](#2,#3)(#4,#5){\@drawline[#1](#2,#3)(#4,#5)
\@idrawline[#1](#4,#5)}
\def\@drawline[#1](#2,#3)(#4,#5){{%
\x@diff=#4\unitlength \advance\x@diff by -#2\unitlength
\y@diff=#5\unitlength \advance\y@diff by -#3\unitlength
\ifx\@linefnt\tenln \linethickness{0.5pt} \else \linethickness{0.9pt}\fi
\lineslope(\x@diff,\y@diff)% returns the two integers in \@xarg & \@yarg.
\@toosmalltrue
{\ifdim\x@diff <\z@ \x@diff=-\x@diff\fi
 \ifdim\y@diff <\z@ \y@diff=-\y@diff\fi
 \ifdim\x@diff >10pt \global\@toosmallfalse\fi
 \ifdim\y@diff >10pt \global\@toosmallfalse\fi}
\@drawitfalse\@horvlinefalse
\ifnum#1 <0 \relax\else\@horvlinetrue\fi
\if@toosmall\@horvlinetrue\fi% to get 'or' condition. We necessarily draw a
\if@horvline
 \ifdim\x@diff =0pt \put(#2,#3){\ifdim\y@diff >0pt \@linelen=\y@diff \@upline
 				\else\@linelen=-\y@diff \@downline\fi}%
 \else\ifdim\y@diff =0pt
          \ifdim\x@diff >0pt \put(#2,#3){\vrule \@height \@halfwidth \@depth
				\@halfwidth \@width \x@diff}
		\else \put(#4,#5){\vrule \@height \@halfwidth \@depth
				\@halfwidth \@width -\x@diff}\fi
       \else\@drawittrue\fi\fi % construct the line explicitly
\else\@drawittrue\fi
\if@drawit
\ifnum\@xarg< 0 \@negargtrue\else\@negargfalse\fi
\ifnum\@xarg =0 \setbox\@linechar%
\hbox{\hskip -\@halfwidth \vrule \@width \@wholewidth \@height 10.2pt
 \@depth \z@}
\else \ifnum\@yarg =0 \setbox\@linechar%
\hbox{\vrule \@height \@halfwidth \@depth \@halfwidth \@width 10.2pt}
\else \if@negarg \@xarg -\@xarg \@yyarg -\@yarg
        \else \@yyarg \@yarg\fi
\ifnum\@yyarg >0 \@tempcnta\@yyarg \else \@tempcnta -\@yyarg\fi
\setbox\@linechar\hbox{\@linefnt\@getlinechar(\@xarg,\@yyarg)}%
\fi\fi
\if@toosmall% => it isn't a horiz or vert line and is toosmall.
  \@dottedline[\picsquare]{.98\@wholewidth}%
(#2\unitlength,#3\unitlength)(#4\unitlength,#5\unitlength)%
\else
\ifnum\@xarg=0\relax\else\ifdim\x@diff >\z@ \advance\x@diff -\wd\@linechar
  \else\advance\x@diff \wd\@linechar\fi\fi
\ifnum\@yarg=0\relax\else\ifdim\y@diff >\z@\advance\y@diff -\ht\@linechar
  \else\advance\y@diff \ht\@linechar\fi\fi
\ifdim\x@diff <\z@ \@x@diff=-\x@diff \else\@x@diff=\x@diff\fi
\ifdim\y@diff <\z@ \@y@diff=-\y@diff \else\@y@diff=\y@diff\fi
\num@segments=0 \num@segmentsi=0
\ifdim\wd\@linechar >1pt
 \num@segmentsi=\@x@diff \divide\num@segmentsi \wd\@linechar\fi
\ifdim\ht\@linechar >1pt
 \num@segments=\@y@diff \divide\num@segments \ht\@linechar\fi
\ifnum\num@segmentsi >\num@segments \num@segments=\num@segmentsi\fi
\advance\num@segments \@ne %to account for round-off error
\ifnum #1=0 \relax \else\ifnum #1 < -99
  \typeout{***drawline: reduction <= -100 percent implies blankness!***}
\else\num@segmentsi=#1 \advance\num@segmentsi by 100
     \multiply\num@segments \num@segmentsi
     \divide\num@segments by 100
\fi\fi
\divide\x@diff \num@segments
\divide\y@diff \num@segments
\advance\num@segments \@ne %for the last segment for which I subtracted
\@xdim=#2\unitlength \@ydim=#3\unitlength
\if@negarg \advance\@xdim -\wd\@linechar\fi
\ifnum\@yarg <0 \advance\@ydim -\ht\@linechar\fi
\@killglue
\loop \ifnum\num@segments > 0
\unskip\raise\@ydim\hbox to\z@{\hskip\@xdim \copy\@linechar\hss}%
\advance\num@segments \m@ne\advance\@xdim\x@diff\advance\@ydim\y@diff%
\repeat
\ignorespaces
\fi%the if of @toosmall
\fi}}% for \if@drawit
\long\def\splittwoargs#1 #2 {(#1,#2)}
\newif\if@stillmore
\newread\@datafile
\long\def\putfile#1#2{\openin\@datafile = #1
\@stillmoretrue
\loop
\ifeof\@datafile\relax\else\read\@datafile to\@dataline\fi
\ifeof\@datafile\@stillmorefalse
\else\ifx\@dataline\@empty \relax
     \else
\expandafter\expandafter\expandafter\put\expandafter\splittwoargs%
\@dataline{#2}
     \fi
\fi
\if@stillmore
\repeat
\closein\@datafile
}
\makeatletter
\newcount\@gphlinewidth
\newcount\@eepictcnt
\newdimen\@tempdimc
\@gphlinewidth\@wholewidth \divide\@gphlinewidth 4736

\def\thinlines{\let\@linefnt\tenln \let\@circlefnt\tencirc
    \@wholewidth\fontdimen8\tenln \@halfwidth .5\@wholewidth
    \@gphlinewidth\@wholewidth \divide\@gphlinewidth 4736\relax}
\def\thicklines{\let\@linefnt\tenlnw \let\@circlefnt\tencircw
    \@wholewidth\fontdimen8\tenlnw \@halfwidth .5\@wholewidth
    \@gphlinewidth\@wholewidth \divide\@gphlinewidth 4736
    \advance\@gphlinewidth\@ne   % Make the output looks better
    \relax}
\newif\if@nodotdef \global\@nodotdeftrue
\def\dottedjoin{\global\@jointhemtrue \global\@joinkind=0\relax
  \bgroup\@ifnextchar[{\global\@nodotdeffalse\@idottedjoin}%
                      {\global\@nodotdeftrue\@idottedjoin[\@empty]}}
\long\def\jput(#1,#2)#3{\@killglue\raise#2\unitlength\hbox to \z@{\hskip
#1\unitlength #3\hss}%
\if@jointhem \if@firstpoint \gdef\x@one{#1} \gdef\y@one{#2}
\global\@firstpointfalse
 \else\ifcase\@joinkind
    \if@nodotdef
        \@spdottedline{\dotgap@join\unitlength}%
(\x@one\unitlength ,\y@one\unitlength)(#1\unitlength,#2\unitlength)
    \else
	\@dottedline[\dotchar@join]{\dotgap@join\unitlength}%
(\x@one\unitlength,\y@one\unitlength)(#1\unitlength,#2\unitlength)
    \fi
	\or\@dashline[\dashlinestretch]{\dashlen@join\unitlength}[\dotgap@join]%
(\x@one,\y@one)(#1,#2)
	\else\@drawline[\drawlinestretch](\x@one,\y@one)(#1,#2)\fi
    \gdef\x@one{#1}\gdef\y@one{#2}%
 \fi
\fi\ignorespaces}
\def\dottedline{\@ifnextchar [{\@idottedline}{\@ispdottedline}}
\def\@ispdottedline#1(#2,#3){\@ifnextchar (%
{\@iispdottedline{#1}(#2,#3)}{\relax}}
\def\@iispdottedline#1(#2,#3)(#4,#5){\@spdottedline{#1\unitlength}%
(#2\unitlength,#3\unitlength)(#4\unitlength,#5\unitlength)%
\@ispdottedline{#1}(#4,#5)}
\def\@spdottedline#1(#2,#3)(#4,#5){%
    \@tempcnta \@gphlinewidth\relax
    \advance\@tempcnta by 2     % solely for better output
    \special{pn \the\@tempcnta}%
    \@tempdimc=#2\relax
    \@tempcnta \@tempdimc\relax \advance\@tempcnta 2368 \divide\@tempcnta 4736
    \@tempdimc=#3\relax
    \@tempcntb -\@tempdimc\relax\advance\@tempcntb -2368 \divide\@tempcntb 4736
    \@paspecial{\the\@tempcnta}{\the\@tempcntb}%
    \@tempdimc=#4\relax
    \@tempcnta \@tempdimc\relax \advance\@tempcnta 2368 \divide\@tempcnta 4736
    \@tempdimc=#5\relax
    \@tempcntb -\@tempdimc\relax\advance\@tempcntb -2368 \divide\@tempcntb 4736
    \@paspecial{\the\@tempcnta}{\the\@tempcntb}%
    \@tempdimc=#1\relax
    \@tempcnta \@tempdimc\relax \advance\@tempcnta 2368 \divide\@tempcnta 4736
    \@tempcntb \@tempcnta\relax \divide\@tempcntb 1000
    \multiply \@tempcntb 1000 \advance\@tempcnta -\@tempcntb
    \divide\@tempcntb 1000
    \ifnum\@tempcnta < 10
        \special{dt \the\@tempcntb.00\the\@tempcnta}%
    \else\ifnum\@tempcnta < 100
        \special{dt \the\@tempcntb.0\the\@tempcnta}%
    \else
        \special{dt \the\@tempcntb.\the\@tempcnta}%
    \fi\fi
    \ignorespaces
}
\def\@iiidashline[#1]#2[#3](#4,#5)(#6,#7){%
\@dashline[#1]{#2\unitlength}[#3](#4,#5)(#6,#7)%
\@iidashline[#1]{#2}[#3](#6,#7)}
\long\def\@dashline[#1]#2[#3](#4,#5)(#6,#7){{%
\x@diff=#6\unitlength \advance\x@diff by -#4\unitlength
\y@diff=#7\unitlength \advance\y@diff by -#5\unitlength
\@tempdima=#2\relax \advance\@tempdima -\@wholewidth
\sqrtandstuff{\x@diff}{\y@diff}{\@tempdima}%
\ifnum\num@segments <3 \num@segments=3\fi% min number of dashes I can plot
\@tempdima=\x@diff \@tempdimb=\y@diff
\divide\@tempdimb by\num@segments
\divide\@tempdima by\num@segments
{\ifx#3\@empty \relax
    \ifdim\@tempdima < 0pt \x@diff=-\@tempdima\else\x@diff=\@tempdima\fi
    \ifdim\@tempdimb < 0pt \y@diff=-\@tempdimb\else\y@diff=\@tempdimb\fi
    \global\setbox\@dotbox\hbox{%
                \@absspdrawline(0pt,0pt)(\@tempdima,\@tempdimb)}%
    \else\global\setbox\@dotbox\hbox{%
        \@spdottedline{#3\unitlength}(0pt,0pt)(\@tempdima,\@tempdimb)}%
    \fi}%
\advance\x@diff by -\@tempdima % both have same sign
\advance\y@diff by -\@tempdimb
\@tempdima=\num@segments\@wholewidth \@tempdima=2\@tempdima
\@tempcnta\@tempdima\relax \@tempdima=#2\relax \@tempdimb=0.5\@tempdima
\@tempcntb\@tempdimb\relax \advance\@tempcnta by \@tempcntb % round-off error
\divide\@tempcnta by\@tempdima \advance\num@segments by -\@tempcnta
\ifnum #1=0 \relax\else\ifnum #1 < -100
  \typeout{***dashline: reduction > -100 percent implies blankness!***}
\else\num@segmentsi=#1 \advance\num@segmentsi by 100
     \multiply\num@segments by\num@segmentsi \divide\num@segments by 100
\fi\fi
\divide\num@segments by 2 % earlier num@segments included 'empty dashes' too.
\ifnum\num@segments >0 % if =0 then don't divide => \x@diff & \y@diff
 \divide\x@diff by\num@segments%   remain same.
 \divide\y@diff by\num@segments
 \advance\num@segments by\@ne %for the last segment for which I subtracted
 \else\num@segments=2 % one at each end.
\fi
\@xdim=#4\unitlength \@ydim=#5\unitlength
\@killglue
\loop \ifnum\num@segments > 0
\unskip\raise\@ydim\hbox to\z@{\hskip\@xdim \copy\@dotbox\hss}%
\advance\num@segments \m@ne\advance\@xdim\x@diff\advance\@ydim\y@diff%
\repeat}%
\ignorespaces}
\def\@drawline[#1](#2,#3)(#4,#5){{%
\@drawitfalse\@horvlinefalse
\ifnum#1 <0 \relax\else\@horvlinetrue\fi
\if@horvline
 \@spdrawline(#2,#3)(#4,#5)
\else\@drawittrue\fi
\if@drawit
\ifnum #1=0 \relax \else\ifnum #1 < -99
  \typeout{***drawline: reduction <= -100 percent implies blankness!***}%
\else\num@segmentsi=#1 \advance\num@segmentsi by 50
     \multiply\num@segmentsi 2
\fi\fi
\@dashline[\num@segmentsi]{10pt}[\@empty](#2,#3)(#4,#5)
\fi}\ignorespaces}% for \if@drawit
\def\@spdrawline(#1,#2)(#3,#4){%
   \@absspdrawline(#1\unitlength,#2\unitlength)(#3\unitlength,#4\unitlength)
   \ignorespaces
}
\def\@absspdrawline(#1,#2)(#3,#4){%
    \special{pn \the\@gphlinewidth}%
    \@tempdimc=#1\relax
    \@tempcnta \@tempdimc\relax \advance\@tempcnta 2368 \divide\@tempcnta 4736
    \@tempdimc=#2\relax
    \@tempcntb -\@tempdimc\relax \advance\@tempcntb -2368 \divide\@tempcntb
4736
    \@paspecial{\the\@tempcnta}{\the\@tempcntb}%
    \@tempdimc=#3\relax
    \@tempcnta\@tempdimc\relax \advance\@tempcnta 2368 \divide\@tempcnta 4736
    \@tempdimc=#4\relax
    \@tempcntb -\@tempdimc\relax \advance\@tempcntb -2368 \divide\@tempcntb
4736
    \@paspecial{\the\@tempcnta}{\the\@tempcntb}%
    \special{fp}%
    \ignorespaces
}
\def\@paspecial#1#2{%
    \special{pa #1 #2}%
}
\def\Thicklines{\let\@linefnt\tenlnw \let\@circlefnt\tencircw
    \@wholewidth\fontdimen8\tenlnw \@wholewidth 1.5\@wholewidth
    \@halfwidth .5\@wholewidth
    \@gphlinewidth\@wholewidth \divide\@gphlinewidth 4736\relax}
\def\@circlespecial#1#2#3#4{%
	      \special{pn \the\@gphlinewidth}%
	      \special{ar 0 0 #1 #2 #3 #4}
}
\def\@arc#1#2#3#4{%
	\@tempdima #1\unitlength
	\@tempdimb #2\unitlength
        \@tempcnta\@tempdima \advance\@tempcnta 4736 \divide\@tempcnta 9473
	\@tempcntb\@tempdimb \advance\@tempcntb 4736 \divide\@tempcntb 9473
	\setbox\@tempboxa\hbox{%
	    \@circlespecial{\the\@tempcnta}{\the\@tempcntb}{#3}{#4}}%
        \wd\@tempboxa\z@ \box\@tempboxa}
\def\circle{%
    \@ifstar{\special{bk}\@circle}{\@circle}}
\def\@circle#1{\@arc{#1}{#1}{0}{6.2832}}
\def\ellipse{%
    \@ifstar{\special{bk}\@ellipse}{\@ellipse}}
\def\@ellipse#1#2{{\@arc{#1}{#2}{0}{6.2832}}}
\def\arc#1#2#3{\@arc{#1}{#1}{#2}{#3}}
\def\@linespecial#1#2{%
	      \special{pn \the\@gphlinewidth}%
	      \special{pa 0 0}%
	      \special{pa #1 #2}%
	      \special{fp}%
}
\def\@sline{%
	\ifnum\@xarg< 0
	  \@negargtrue \@xarg -\@xarg \@tempdima -\@linelen
	\else
	  \@negargfalse \@tempdima\@linelen
	\fi
	\@tempcnta\@linelen \divide\@tempcnta 4736
        \@yyarg -\@yarg \multiply\@yyarg \@tempcnta \divide\@yyarg\@xarg
 	\if@negarg
	    \@tempcnta -\@tempcnta
	\fi
	\setbox\@linechar\hbox{\@linespecial{\the\@tempcnta}{\the\@yyarg}}%
	\wd\@linechar\@tempdima
	\@clnht\@linelen
        \multiply\@clnht\@yarg
        \divide\@clnht\@xarg
	\ifnum\@yarg< 0
	  \@clnht -\@clnht
	  \ht\@linechar\z@ \dp\@linechar\@clnht
	\else
	  \ht\@linechar\@clnht \dp\@linechar\z@
	\fi
	\box\@linechar
	\if@negarg
	  \@yyarg -\@yarg
	\else
	  \@yyarg \@yarg
	\fi
	\setbox\@linechar\hbox{\@linefnt\@getlinechar(\@xarg,\@yyarg)}%
	\ifnum\@yarg> 0
	  \let\@upordown\raise
	  \advance\@clnht -\ht\@linechar
	\else
	  \let\@upordown\lower
	\fi
	\if@negarg \kern\wd\@linechar \fi
}
\def\spline(#1,#2){%
    \special{pn \the\@gphlinewidth}%
    \@spline(#1,#2)%
}
\def\@spline(#1,#2){%
    \@tempdima #1\unitlength
    \@tempdimb #2\unitlength
    \@tempcnta \@tempdima \advance\@tempcnta 2368 \divide\@tempcnta 4736
    \@tempcntb -\@tempdimb \advance\@tempcntb -2368 \divide\@tempcntb 4736
    \@paspecial{\the\@tempcnta}{\the\@tempcntb}%
    \@ifnextchar ({\@spline}{\special{sp}}%
}
\def\path(#1,#2){%
    \special{pn \the\@gphlinewidth}%
    \@path(#1,#2)%
}
\def\@path(#1,#2){%
    \@tempdima #1\unitlength
    \@tempdimb #2\unitlength
    \@tempcnta \@tempdima \advance\@tempcnta 2368 \divide\@tempcnta 4736
    \@tempcntb -\@tempdimb \advance\@tempcntb -2368 \divide\@tempcntb 4736
    \@paspecial{\the\@tempcnta}{\the\@tempcntb}%
    \@ifnextchar ({\@path}{\special{fp}}%
}

\newdimen\maxovaldiam \maxovaldiam 40pt\relax

\def\@oval(#1,#2)[#3]{\begingroup\boxmaxdepth \maxdimen
  \@ovttrue \@ovbtrue \@ovltrue \@ovrtrue
  \@tfor\@tempa :=#3\do{\csname @ov\@tempa false\endcsname}\@ovxx
  #1\unitlength \@ovyy #2\unitlength
  \@tempdimb \ifdim \@ovyy >\@ovxx \@ovxx\else \@ovyy \fi
  \@ovro \ifdim \@tempdimb>\maxovaldiam \maxovaldiam\else\@tempdimb\fi\relax
  \divide \@ovro \tw@
  \@ovdx\@ovxx \divide\@ovdx \tw@
  \@ovdy\@ovyy \divide\@ovdy \tw@
  \setbox\@tempboxa \hbox{%
  \if@ovr \@ovverta\fi
  \if@ovl \kern \@ovxx \@ovvertb\kern -\@ovxx \fi
  \if@ovt \@ovhorz \kern -\@ovxx \fi
  \if@ovb \raise \@ovyy \@ovhorz \fi}% bug found by isozaki
  \ht\@tempboxa\z@ \dp\@tempboxa\z@
  \@put{-\@ovdx}{-\@ovdy}{\box\@tempboxa}%
  \endgroup}

\def\@qcirc#1#2#3#4{%
    \special{pn \the\@gphlinewidth}%
    \@eepictcnt \@gphlinewidth \divide\@eepictcnt 2
    \@tempcnta #1
      \advance\@tempcnta 2368 \divide\@tempcnta 4736
      \advance\@tempcnta\@eepictcnt
    \@tempcntb #2 \divide\@tempcntb 4736 \advance\@tempcntb 2
    \hbox{%
\@qcircspecial{\the\@tempcnta}{-\the\@eepictcnt}{\the\@tempcntb}{#3}{#4}}%
}
\def\@qcircspecial#1#2#3#4#5{\special{ar #1 #2 #3 #3 #4 #5}}

\def\@ovverta{\vbox to \@ovyy{%
    \if@ovb
        \kern \@ovro
        \@qcirc{\@ovro}{\@ovro}{3.1416}{4.7124}\nointerlineskip
    \else
        \kern \@ovdy
    \fi
    \leaders\vrule width \@wholewidth\vfil \nointerlineskip
    \if@ovt
        \@qcirc{\@ovro}{\@ovro}{1.5708}{3.1416}\nointerlineskip
        \kern \@ovro
    \else
        \kern \@ovdy
    \fi
}\kern -\@wholewidth}

\def\@ovvertb{\vbox to \@ovyy{%
    \if@ovb
        \kern \@ovro
        \@qcirc{-\@ovro}{\@ovro}{4.6124}{6.2832}\nointerlineskip
    \else
        \kern \@ovdy
    \fi
    \leaders\vrule width \@wholewidth\vfil \nointerlineskip
    \if@ovt
        \@qcirc{-\@ovro}{\@ovro}{0}{1.6708}\nointerlineskip
        \kern \@ovro
    \else
        \kern \@ovdy
    \fi
}\kern -\@wholewidth}

\def\@ovhorz{\hbox to \@ovxx{%
    \if@ovr \kern \@ovro\else \kern \@ovdx \fi
    \leaders \hrule height \@wholewidth \hfil
    \if@ovl \kern \@ovro\else \kern \@ovdx \fi
    }}
\def\allinethickness#1{\let\@linefnt\tenlnw \let\@circlefnt\tencircw
    \@wholewidth #1 \@halfwidth .5\@wholewidth
    \@gphlinewidth\@wholewidth \divide\@gphlinewidth 4736\relax}
\title{Difference equations for the correlation functions  \\
of the eight-vertex model}
\author{
Michio Jimbo\thanks{
Department of Mathematics, Faculty of Science,
Kyoto University, Kyoto 606, Japan}
\and
Tetsuji Miwa\thanks{
Research Institute for Mathematical Sciences,
Kyoto University, Kyoto 606, Japan}
\and
Atsushi Nakayashiki\thanks{
The Graduate School of Science and Technology,
Kobe University, Rokkodai, Kobe 657, Japan}
}
\date{December 6, 1992}

\begin{document}

\maketitle
\begin{abstract}
We propose that the correlation functions of the inhomogeneous
eight-vertex model in the anti-ferroelectric regime satisfy a system of
difference equations with respect to the spectral parameters.
Solving the simplest difference equation
we obtain the expression for the  spontaneous staggered
polarization conjectured by Baxter and Kelland.
We also discuss a related construction of vertex operators on the lattice.
\end{abstract}

\setcounter{section}{-1}
\setcounter{secnumdepth}{1}
\def\C{{\bf C}}
\def\R{{\bf R}}
\def\Rb{\overline{R}}
\def\Rc{\check R}
\def\Q{{\bf Q}}
\def\Z{{\bf Z}}
\def\Zp{{\bf Z}_{\ge 0}}
\def\Zm{{\bf Z}_{\le 0}}
\def\A{{\cal A}}
\def\H{{\cal H}}
\def\Ca{{\cal C}}
\def\F{{\cal F}}
\def\Fr{{\F}^r}
\def\Y{{\cal Y}}
\def\S{{\bf S}}
\def\P{{\cal P}}
\def\eL{{\cal L}}
\def\la{\lambda}
\def\La{\Lambda}
\def\Ga{\Gamma}
\def\Laz{u_{\La_0}}
\def\Lao{u_{\La_1}}
\def\Th{\Theta}
\def\th{\theta}
\def\Et{H}
\def\zt{\zeta}
\def\qb{\bar{q}}
\def\rb{\bar{r}}
\def\Vh{\widehat{V}}
\def\Vt{\widetilde{V}}
\def\Lh{\widehat{L}}
\def\vep{\varepsilon}
\def\vph{\varphi}
\def\ep{\epsilon}
\def\mod{\hbox{mod}}
\def\Hom{\hbox{Hom}}
\def\End{\hbox{End}}
\def\id{\hbox{id}}
\def\tr{\hbox{tr}}
\def\Tr{\hbox{Tr}}
\def\deg{\hbox{deg}}
\def\wt{\hbox{wt}\,}
\def\gl{\goth{gl}_N}
\def\slt{\goth{sl}_2}
\def\slth{\widehat{\goth{sl}}_2\hskip 2pt}
\def\uq{U_q(\goth{g})}
\def\uqa{U_q(\slth)}
\def\goto#1{{\buildrel #1 \over \longrightarrow}}
\def\br#1{\langle #1 \rangle}
\def\bra#1{\langle #1 |}
\def\ket#1{|#1\rangle}
\def\brak#1#2{\langle #1|#2\rangle}
\def\chk#1{#1^\vee}
\def\vac{|\hbox{vac}\rangle}
\def\dvac{\langle \hbox{vac}|}
\def\Phit{\widetilde{\Phi}}
\def\Psit{\widetilde{\Psi}}
\def\ft{\tilde{f}}
\def\et{\tilde{e}}
\def\fti{\tilde{f}_i}
\def\eti{\tilde{e}_i}
%
%\goth
\font\germ=eufm10
\def\goth#1{\hbox{\germ #1}}
\def\sectiontitle#1\par{\vskip0pt plus.1\vsize\penalty-250
 \vskip0pt plus-.1\vsize\bigskip\vskip\parskip
 \message{#1}\leftline{\bf#1}\nobreak\vglue 5pt}
\def\qed{\hbox{${\vcenter{\vbox{
    \hrule height 0.4pt\hbox{\vrule width 0.4pt height 6pt
    \kern5pt\vrule width 0.4pt}\hrule height 0.4pt}}}$}}
\def\subsec(#1|#2){\medskip\noindent{\it #1}\hskip8pt{\it #2}\quad}
\def\nnb{\nonumber}
\def\eqa#1\endeqa
	{\begin{eqnarray}
		#1
	\end{eqnarray}}
\def\eq#1\endeq
	{\begin{eqnarray*}
		#1
	\end{eqnarray*}}
\def\refeq#1{(\ref{eqn:#1})}
\def\refto#1{\cite{#1}}
%
%\subsec
%
\def\subsec(#1|#2){\medskip\noindent#1\hskip8pt{\sl #2}\quad}
%
%\qed
%
%\inpu box
%\def\qed{\qquad$\Fsquare(.2cm,{})$}
%
%\abstract
%
\def\abstract#1\endabstract{
\bigskip
\itemitem{{}}
{\bf Abstract.}
\quad
#1
\bigskip
}
%
%
%\section
%
\def\sec(#1){Sect.\hskip2pt#1}

\section{Introduction}

In \cite{DFJMN,JMMN} it was recognized
that the correlation functions of the inhomogeneous six-vertex model
in the anti-ferroelectric regime can be expressed as a trace of products of
the $q$-deformed vertex operators. An explicit integral formula is given in
\cite{JMMN}. These correlators satisfy a system of
$q$-difference equations \cite{IIJMNT},
that were introduced by Smirnov in the study of
form factors in massive integrable QFT \cite{Smbk}
and correspond to the level-$0$ case of the $q$-KZ equation
of Frenkel and Reshetikhin \cite{FR}.
(To be precise
the equations for the correlators are `dual' to Smirnov's ones, but
we do not go into such details here).
As remarked in \cite{FR},
if one replaces the trigonometric $R$ matrices appearing here by
the elliptic ones, the resulting equations are still `completely integrable'.
In this paper we propose that the latter are precisely those
satisfied by the correlators
of the eight-vertex model in the anti-ferroelectric regime.

Our method is based simply
on the Yang-Baxter equation and the crossing symmetry and, as such,
is applicable to more general models.
This construction also allows one to interpret the $q$-vertex operator
employed in \cite{DFJMN,JMMN}
as an operator that inserts a dislocation (an extra half infinite line)
on the lattice.
In this formulation the vertex operators
generalize straightforwardly to the elliptic case.
It remains an interesting open problem to give a
mathematical construction of such operators, along with the relevant
elliptic deformation of Lie algebras.

The text is organized as follows.
In \sec(1), we introduce our notation, and formulate the properties
of the general correlators including Smirnov's difference equations.
We give a heuristic argument for derivation of the main statements.
In \sec(2) we solve the simplest difference equation and
derive the spontaneous staggered polarization.
We obtain the expression conjectured by Baxter and Kelland \cite{BK}.
Sect.3 is devoted to the construction of vertex operators
on the lattice.
\def\R{{\cal R}}
\def\ve{\varepsilon}
\def\kab{\bar{\kappa}}
\def\sub#1|#2|{\smallskip\noindent{\sl #1\hskip5pt #2\qquad}}
\def\ip(#1;#2){(#1;#2)_\infty}
\def\iq(#1){(#1;q,x^4)_\infty}
%\input fin1.tex
% Miwa-sensei,
%      Maybe this works well, but if #6 is long it doesn't.
% As you know I am not so good at making TeX macros.
% Our TeX teachers are only YOU and Nakahara-san who are
% too busy to teach us something.
%     If I had time, I would rather rewrite this by using
% \matrix or something.       K.S.
%
\def\bp(#1,#2,#3,#4,#5,#6){
\dimen6=1.1cm
\dimen7=\dimen6
\dimen10=\dimen6
\advance \dimen10 by 6pt
\divide \dimen6 by 2
\dimen8=\dimen6
\dimen9=\dimen6
\advance \dimen8 by 0.2pt
\advance \dimen9 by -0.2pt
\dimen11=\dimen6
\divide \dimen11 by 2
\advance \dimen11 by -.02cm
\dimen16=\dimen6
\advance \dimen16 -0.09cm
#4
\raise1.8pt\hbox{$
{\mathop{\hbox to \dimen10{
\vrule width \dimen6 depth0pt height.3pt \vrule width \dimen11 depth0pt
    height.3pt
\hskip-.2mm
\lower.799mm\hbox{$\leftarrow$}
\hskip-.77cm    %\dimen13
\raise3.2mm\hbox{$\downarrow$}
\hskip-.215cm
\vrule height \dimen11 depth \dimen6 width.3pt
\hskip0.1cm
\raise.2cm\hbox to 0.31cm{$#5$ \hss}
\hskip-.35cm
\lower.33cm\hbox to 0.31cm{\hss $#6$}
\hfill}}
\limits^{\textstyle \; #1}_{\textstyle \; #3}}$}
#2
}
\def\bpup(#1,#2,#3,#4,#5,#6){
\dimen6=1.1cm
\dimen7=\dimen6
\dimen10=\dimen6
\advance \dimen10 by 6pt
\divide \dimen6 by 2
\dimen8=\dimen6
\dimen9=\dimen6
\advance \dimen8 by 0.2pt
\advance \dimen9 by -0.2pt
\dimen11=\dimen6
\divide \dimen11 by 2
\advance \dimen11 by -0.02cm
\setbox12=\hbox{$\leftarrow$}
\dimen12=\wd12
\dimen13=\dimen12
\advance \dimen13 by \dimen8
\setbox14=\hbox{$\downarrow$}
\dimen14=\wd14
\dimen15=\dimen14
\divide \dimen15 by 2
\setbox16=\hbox{$\uparrow$}
#4
\raise1.8pt\hbox{$
{\mathop{\hbox to \dimen10{
%\vrule width \dimen6 depth0pt height.3pt \vrule width \dimen6 depth0pt
%    height.3pt
%\hskip-.55cm    %\dimen13
\vrule width \dimen6 depth0pt height.3pt \vrule width \dimen11 depth0pt
    height.3pt
\hskip-.2mm
\lower.799mm\hbox{$\leftarrow$}
\hskip-.72cm    %\dimen13
\raise3.2mm\hbox{%
\vrule height \dimen11 depth0pt width.3pt}
\hskip-.21cm
\raise.85mm\hbox{\copy16}
\hskip-.215cm
\vrule height0pt depth \dimen6 width.3pt
%\hskip \dimen9
%\hskip-.64cm
\hskip0.1cm
\raise.2cm\hbox to 0.31cm{$#5$\hss}
\hskip-.38cm
\lower.33cm\hbox to 0.31cm{\hss $#6$}
\hfill}}
\limits^{\textstyle #1}_{\textstyle #3}}$}
#2
}

\section{Correlators and Smirnov's equations}

\sub1.1|The eight-vertex model|
Consider an infinite square lattice consisting of oriented lines,
each carrying a spectral parameter varying from line to line.
The orientation of each line will be shown by an arrow on it.
A vertex is a crossing of two lines with spectral parameters,
say $\zeta_1$ and $\zeta_2$, together with the adjacent 4
edges belonging to the crossing lines, as shown in \refeq{vertex}.
The edges are assigned state variables (which we call `spins')
: $\ve_1,\ve_2,\ve'_1,\ve'_2$. In the eight-vertex model, each
spin can take one of the two different values $\pm$.
A spin configuration around the vertex is an
assignment of $\pm$ on the four edges.
Notice that our notation is somewhat different from the standard.
In particular, arrows on the lines denote the orientations that we assign
to them, rather than the spins.
There are 16 possible vertex configurations.
We assign each configuration a Boltzmann weight.
The set of all Boltzmann weights form the elements of the $R$-matrix:
\eqa
&&\bp(\ve_1,\ve_2,\ve'_1,\ve'_2,\zeta_1,\zeta_2)=
R(\zeta_1/\zeta_2)_{\ve_1'\,\ve_2';\ve_1\,\ve_2}. \label{eqn:vertex}
\endeqa
The matrix $R(\zeta)$ acts on $\C^2\otimes\C^2$
via the natural basis $\{v_+, v_-\}$ of $\C^2$ as
$R(\zeta)v_{\ve_1'}\otimes v_{\ve_2'}
=\sum v_{\ve_1}\otimes v_{\ve_2} R(\zeta)_{\ve_1\,\ve_2\,;\,\ve_1'\,\ve_2'}$.
Normalized by the partition function per site $\kappa$,
it is given explicitly by
\eq
R(\zeta)=
{1\over\kappa(\zeta)}\bordermatrix{&++&+-&-+&--\cr
++&a(\zeta)&0&0&d(\zeta)\cr
+-&0&b(\zeta)&c(\zeta)&0\cr
-+&0&c(\zeta)&b(\zeta)&0\cr
--&d(\zeta)&0&0&a(\zeta)\cr}
\endeq
where the unnormalized Boltzmann weights
$a,b,c,d$ (see (10.4.23-24) and (10.7.9) in \cite{Baxbk})
are given by
\eq
a(\zeta)&=-i\rho\Theta(i\lambda)H(i\lambda-iu)\Theta(iu),\quad
b(\zeta)&=-i\rho\Theta(i\lambda)\Theta(i\lambda-iu)H(iu),\\
c(\zeta)&=-i\rho H(i\lambda)\Theta(i\lambda-iu)\Theta(iu),\quad
d(\zeta)&=i\rho H(i\lambda)H(i\lambda-iu)H(iu),
\endeq
where we set
$q=exp(-\pi I'/I)$, $x=exp(-\pi \lambda/2I)$, $\zeta=exp(\pi u/2I)$.
We shall restrict our discussion to the principal regime
\eq
&& 0 < q < x < \zeta^{-1} < 1,
\endeq
in which $\kappa(\zeta)$
is given by (see (10.8.44), ibid.)
\eq
\kappa(\zeta)&=&
\rho\gamma x^{-1} \kab(\zeta^2)
\ip(x^2\zeta^2;q)\ip(qx^{-2}\zeta^{-2};q),
\quad\gamma=q^{1/4}(q;q)_\infty^2(q^2;q^2)_\infty,\\
\kab(z)&=&
{\iq(x^4z)\iq(x^2z^{-1})\iq(qz)\iq(qx^2z^{-1})
\over
\iq(x^4z^{-1})\iq(x^2z)\iq(qz^{-1})\iq(qx^2z)}.
\endeq
The symbol $\ip(z;q_1,\ldots,q_k)$ means
$\prod_{n_1,\cdots,n_k=0}^\infty (1-zq_1^{n_1}\cdots q_k^{n_k})$.
Notice that in the definition of the $R$-matrix,
$\kappa$ was used to normalize the Boltzmann weights.

The $R$-matrix satisfies
\eqa
&&R(\zeta)PR(\zeta^{-1})P=1,  \nnb \\
&&\left(PR(\zeta^{-1})P\right)^{t_1}=(\sigma^x\otimes 1)R(\zeta/x)
(\sigma^x\otimes 1)^{-1}.\label{eqn:cross}
\endeqa
Here $Pv\otimes v'=v'\otimes v$, $(\cdot)^{t_1}$ means the
transpose with respect to the first component,
and $\sigma^x=\pmatrix{0&1\cr 1&0 \cr}$.
We note that
\eqa
&&R(x^{-1})=\pmatrix{0&&&0\cr&1&1&\cr&1&1&\cr0&&&0\cr}, \label{eqn:Rx}
\endeqa
and that Eq.\refeq{cross} is equivalent to
\eq
&&\bpup(\ve_1,\ve_2,\ve'_1,\ve'_2,\zeta_1,\zeta_2)=
\bp(-\ve_1,\ve_2,-\ve'_1,\ve'_2,\zeta_1',\zeta_2),\qquad\zeta'_1=\zeta_1/x.
\endeq
Namely, reversing the orientation of a vertical line in \refeq{vertex}
gives rise to
reversing the spins on that line and shifting the spectral
parameter $\zeta_1\rightarrow\zeta'_1=\zeta_1/x$.

\sub 1.2|Correlators and dislocations on the lattice|
Let us consider an infinite
square lattice \cite{Roy} with all the vertical lines oriented downward,
all the horizontal lines to the left. We associate the spectral parameters
$\zeta_j$ to the vertical lines, and $\xi_k$ to the horizontal lines.
As argued in \cite{Roy}, the calculation of
correlators of arbitrary spins reduces to calculating correlators of
the vertical-edge spins located on the same row, where by `row' we mean
the set of all vertical edges between two neighboring horizontal lines.
We shall restrict our attention to this case.

In the principal regime we are considering, the Boltzmann weight
$c$ dominates the others.
In the low temperature limit: $q$, $x$, $\xi_k/\zeta_j\rightarrow0$,
only type $c$ vertex-configurations are non-vanishing.
Therefore, the spin variables take the same value in
the NE-SW direction, and alternate in the NW-SE direction.
In this limit, two spin configurations are possible:
the ground state configurations (GSC's).
At finite but low temperatures, we choose and fix a GSC, and consider
the statistical sum over all configurations which differ by a finite number
of spins from that GSC.
We let $i\in \Z_2$ label the choice of GSC by the fact that
in the low temperature limit the spin on some reference edge
is frozen to the value $(-1)^{i+1}$. The choice of the reference
edge will be given later.
Now choose a particular row, and consider $n$ successive vertical
edges on it. Let $\zeta_1,\ldots,\zeta_n$ be the spectral parameters
attached to the corresponding lines, numbered from left to right.
As the reference edge we take the next left to the one
with the spectral parameter $\zeta_1$. Consider the probability of the spins
taking the values $\ve_1,\ldots,\ve_n\in\{\pm\}$.
Baxter's conclusion \cite{Roy} was that this correlator is independent of
all other spectral parameters.

In this work, we consider sums over all spin configurations on
various lattices, with certain spins kept fixed.
On infinite lattices, such configuration sums are not well-defined.
Only the ratios of such sums, on the same lattice but with the
different choices of fixed spins, are well-defined in low temperature
series expansions. When considering probabilities,
the normalization is so chosen
that the sum of all correlators is equal to 1. However,
except for the simplest situation as \refeq{correl},
such a choice does not give solutions to our system of equations.
In fact, we do not know how to normalize our correlators. We
postulate that there exists a proper normalization
such that they satisfy all the required properties given below.

In order to obtain expressions that satisfy our difference equations
we consider more general correlators than the probabilities of spins.
We recover the original spin correlators by specializations
(see \refeq{correl}).
We proceed as follows:
We break the $n$ lines at the chosen row, and change the spectral parameters
of the lower halves to $\zeta'_1,\ldots,\zeta'_n$. Then we consider
the configuration sum with the $2n$ spins fixed to the values
$\ve_1,\ldots,\ve_n$ from left to right on the upper row,
and $\ve'_1,\ldots,\ve'_n$ on the lower (see Figure 1.1).
We also consider a `dislocated' lattice obtained as follows.
Delete the $n$ lower half lines with the spectral parameters
$\zeta'_1,\cdots,\zeta'_n$. Therefore, we are left
with the $n$ upper half lines whose spectral parameters are
$\zeta_1,\ldots,\zeta_n$. Rename them to $\xi_{n+1},\ldots,\xi_{2n}$.
Now insert another set of
$n$ upper half lines with spectral parameters $\xi_1,\ldots,\xi_n$
at the next left to those which already exist.
The orientation of the inserted lines is downward.
Thus we obtain $2n$ half lines with the spectral parameters
$\xi_1,\ldots,\xi_{2n}$ from left to right (see Figure 1.2).
We consider the correlator of the $2n$ spins $\tau_1,\ldots,\tau_{2n}$
on the bottom edges of these half lines,
and denote it by
$F^{(i)}_{2n}(\xi_1,\ldots,\xi_{2n})_{\tau_1,\ldots,\tau_{2n}}$.

%\Fig(f11.tex)
\setlength{\unitlength}{0.0125in}
\begin{picture}(251,301)(0,-10)
\drawline(160,145)(160,65)
\drawline(158.000,73.000)(160.000,65.000)(162.000,73.000)
\drawline(80,145)(80,65)
\drawline(78.000,73.000)(80.000,65.000)(82.000,73.000)
\drawline(160,265)(160,185)
\drawline(158.000,193.000)(160.000,185.000)(162.000,193.000)
\drawline(80,265)(80,185)
\drawline(78.000,193.000)(80.000,185.000)(82.000,193.000)
\drawline(200,265)(200,65)
\drawline(198.000,73.000)(200.000,65.000)(202.000,73.000)
\drawline(40,265)(40,65)
\drawline(38.000,73.000)(40.000,65.000)(42.000,73.000)
\drawline(8.000,107.000)(0.000,105.000)(8.000,103.000)
\drawline(0,105)(240,105)
\drawline(8.000,227.000)(0.000,225.000)(8.000,223.000)
\drawline(0,225)(240,225)
\put(155,155){\makebox(0,0)[lb]{\raisebox{0pt}[0pt][0pt]{\shortstack[l]{{\twlrm
$\varepsilon'_n$}}}}}
\put(75,155){\makebox(0,0)[lb]{\raisebox{0pt}[0pt][0pt]{\shortstack[l]{{\twlrm
$\varepsilon'_1$}}}}}
\put(155,175){\makebox(0,0)[lb]{\raisebox{0pt}[0pt][0pt]{\shortstack[l]{{\twlrm
$\varepsilon_n$}}}}}
\put(75,175){\makebox(0,0)[lb]{\raisebox{0pt}[0pt][0pt]{\shortstack[l]{{\twlrm
$\varepsilon_1$}}}}}
\put(155,50){\makebox(0,0)[lb]{\raisebox{0pt}[0pt][0pt]{\shortstack[l]{{\twlrm
$\zeta'_n$}}}}}
\put(75,50){\makebox(0,0)[lb]{\raisebox{0pt}[0pt][0pt]{\shortstack[l]{{\twlrm
$\zeta'_1$}}}}}
\put(155,275){\makebox(0,0)[lb]{\raisebox{0pt}[0pt][0pt]{\shortstack[l]{{\twlrm
$\zeta_n$}}}}}
\put(75,275){\makebox(0,0)[lb]{\raisebox{0pt}[0pt][0pt]{\shortstack[l]{{\twlrm
$\zeta_1$}}}}}
\put(110,240){\makebox(0,0)[lb]{\raisebox{0pt}[0pt][0pt]{\shortstack[l]{{\twlrm
$\cdots$}}}}}
\put(115,80){\makebox(0,0)[lb]{\raisebox{0pt}[0pt][0pt]{\shortstack[l]{{\twlrm
$\cdots$}}}}}
\put(15,15){\makebox(0,0)[lb]{\raisebox{0pt}[0pt][0pt]{\shortstack[l]{{\twlrm
Figure 1.1\quad Breaking the successive}}}}}
\put(15,0){\makebox(0,0)[lb]{\raisebox{0pt}[0pt][0pt]{\shortstack[l]{{\twlrm
$n$ vertical lines.}}}}}
\end{picture}

One of our postulates is that the correlator of Figure 1.1 is given by
\eq
&&F^{(i+n)}_{2n}(\zeta'_n/x,\ldots,\zeta'_1/x,\zeta_1,\ldots,\zeta_{n})
_{-\ve'_n,\ldots,-\ve'_1,\ve_1,\ldots,\ve_n}.
\endeq
In \sec(1.3) we will give more general statements than this, see \refeq{rot1},
\refeq{rot2}.
If we restrict the variables in the above expression as follows
\eqa
&&F^{(i+n)}_{2n}(\zeta_n/x,\ldots,\zeta_1/x,\zeta_1,\ldots,\zeta_{n})
_{-\ve_n,\ldots,-\ve_1,\ve_1,\ldots,\ve_n}, \label{eqn:correl}
\endeqa
we obtain the probability of the original $n$ spins
taking the values $\ve_1,\ldots,\ve_n$ on the regular infinite lattice
(see \refeq{Nor1}).

Hereafter we will
write $n,\zeta_j,\ve_j$ for $2n,\xi_j,\tau_j$ in $F^{(i)}$,
with the understanding that if $n$ is odd, the correlator is zero. Set
\eq
F^{(i)}_n(\zeta_1,\ldots,\zeta_n)&&=
\sum_{\ve_1,\ldots,\ve_n=\pm}
F^{(i)}_n(\zeta_1,\ldots,\zeta_n)_{\ve_1,\ldots,\ve_n}
v_{\ve_1}\otimes\cdots\otimes v_{\ve_n} \\
&&\qquad\qquad\quad \in \C^2\otimes\cdots\otimes\C^2.
\endeq
We will denote by $R_{jk}(\zeta)\ (j<k)$ the matrix $R(\zeta)$
acting on the $j$-th and the $k$-th tensor components. We also use
the transposition $P_{jk}=R_{jk}(1)$.

%\Fig(f12.tex)
\setlength{\unitlength}{0.0125in}
\begin{picture}(338,241)(0,-10)
\drawline(280,205)(280,45)
\drawline(278.000,53.000)(280.000,45.000)(282.000,53.000)
\drawline(40,205)(40,45)
\drawline(38.000,53.000)(40.000,45.000)(42.000,53.000)
\drawline(200,205)(200,125)
\drawline(198.000,133.000)(200.000,125.000)(202.000,133.000)
\drawline(240,205)(240,125)
\drawline(238.000,133.000)(240.000,125.000)(242.000,133.000)
\drawline(120,205)(120,125)
\drawline(118.000,133.000)(120.000,125.000)(122.000,133.000)
\drawline(80,205)(80,125)
\drawline(78.000,133.000)(80.000,125.000)(82.000,133.000)
\drawline(8.000,87.000)(0.000,85.000)(8.000,83.000)
\drawline(0,85)(320,85)
\drawline(8.000,167.000)(0.000,165.000)(8.000,163.000)
\drawline(0,165)(320,165)
\put(230,110){\makebox(0,0)[lb]{\raisebox{0pt}[0pt][0pt]{\shortstack[l]{{\twlrm
$\tau_{2n}$}}}}}
\put(75,110){\makebox(0,0)[lb]{\raisebox{0pt}[0pt][0pt]{\shortstack[l]{{\twlrm
$\tau_1$}}}}}
\put(235,215){\makebox(0,0)[lb]{\raisebox{0pt}[0pt][0pt]{\shortstack[l]{{\twlrm
$\xi_{2n}$}}}}}
\put(75,215){\makebox(0,0)[lb]{\raisebox{0pt}[0pt][0pt]{\shortstack[l]{{\twlrm
$\xi_1$}}}}}
\put(155,135){\makebox(0,0)[lb]{\raisebox{0pt}[0pt][0pt]{\shortstack[l]{{\twlrm
$\cdots$}}}}}
\put(20,0){\makebox(0,0)[lb]{\raisebox{0pt}[0pt][0pt]{\shortstack[l]{{\twlrm
upper half lines.}}}}}
\put(20,15){\makebox(0,0)[lb]{\raisebox{0pt}[0pt][0pt]{\shortstack[l]{{\twlrm
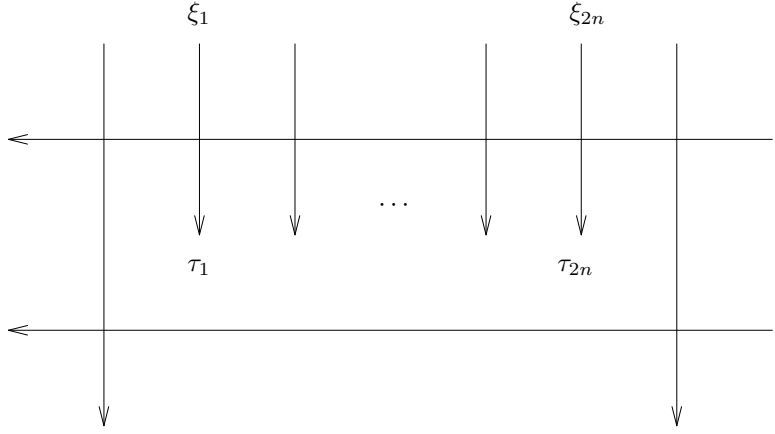
Figure 1.2\quad The dislocated lattice with $2n$}}}}}
\end{picture}

The following are the main results of this paper.
In \sec(1.3) we will give a heuristic proof by
considering
$F^{(i)}_n(\zeta_1,\ldots,\zeta_n)_{\ve_1,\ldots,\ve_n}$
as a sum over all configurations with the spins
$\ve_1,\ldots,\ve_n$ kept fixed.
\proclaim Difference Equation.
\eqa
&&F^{(i)}_n(\zeta_1,\ldots,x^2\zeta_j,\ldots,\zeta_n)
=R_{j\,j+1}(x^2\zeta_j/\zeta_{j+1})^{-1}\cdots
R_{jn}(x^2\zeta_j/\zeta_n)^{-1}  \nnb \\
&&\qquad \times R_{1j}(\zeta_1/\zeta_j)\cdots
R_{j-1j}(\zeta_{j-1}/\zeta_j)
F^{(i+1)}_n(\zeta_1,\ldots,\zeta_j,\ldots,\zeta_n). \label{eqn:Diff}
\endeqa

\proclaim $R$-Matrix Symmetry.
\eqa
&&F^{(i)}_n(\ldots,\zeta_{j+1},\zeta_j,\ldots)
=P_{j\,j+1}R_{j\,j+1}(\zeta_j/\zeta_{j+1})
F^{(i)}_n(\ldots,\zeta_j,\zeta_{j+1},\ldots).\label{eqn:Symm}
\endeqa

\proclaim $\Z_2$-Invariance and Parity.
\eqa
F^{(i+1)}_n(\zeta_1,\ldots,\zeta_n)_{\ve_1,\ldots,\ve_n}
&=&F^{(i)}_n(\zeta_1,\ldots,\zeta_n)_{-\ve_1,\ldots,-\ve_n},\label{eqn:Z2i1}\\
F^{(i)}_n(\ldots,-\zeta_j,\ldots)_{\ldots,\ve_j,\ldots}
&=&(-)^{i+j-1}\ve_j
F^{(i)}_n(\ldots,\zeta_j,\ldots)_{\ldots,\ve_j,\ldots}. \label{eqn:Z2i2}
\endeqa

\proclaim Normalization.
\eqa
&&\sum_{\ve=\pm}F^{(i)}_{n+2}(\ldots,\zeta_{j-1},\zeta,x\zeta,\zeta_{j},\ldots)
_{\ldots,\ve_{j-1},\ve,-\ve,\ve_{j},\ldots} \nnb \\
&&\phantom{\sum_{\ve=\pm}}
=F^{(i)}_n(\ldots,\zeta_{j-1},\zeta_{j},\ldots)
_{\ldots,\ve_{j-1},\ve_{j},\ldots}, \label{eqn:Nor1}\\
&&F^{(i)}_{n+2}(\ldots,\zeta_{j-1},x\zeta,\zeta,\zeta_{j},\ldots)
_{\ldots,\ve_{j-1},\ve,\ve',\ve_{j},\ldots} \nnb\\
&&\phantom{\sum_{\ve=\pm}}
=\delta_{\ve,-\ve'}F^{(i)}_n(\ldots,\zeta_{j-1},\zeta_{j},\ldots)
_{\ldots,\ve_{j-1},\ve_{j},\ldots}. \label{eqn:Nor2}
\endeqa

\sub1.3|Rotating a half line|
Let us consider the following more general correlators;
Choose a face of the lattice. Let $n_1+n_2=n$ (even).
Insert $n_1$ upper half lines and $n_2$ lower half lines
into the face, all of which are oriented downward. Note that in \sec(1.2)
we considered the cases where $n_1=n_2$ or $n_2=0$.
We let the edge on the west side of the said face
to be the reference edge used to label the sectors,
i.e., in the low temperature limit the spin on that edges is
$(-1)^{i+1}$. (See Figure 2.1.) Denote the correlator of this
dislocated lattice by
$F^{(i)}(\zeta_1,\ldots,\zeta_{n_1};\zeta'_1,\ldots,\zeta'_{n_2})
_{\ve_1,\ldots,\ve_{n_1};\ve'_1,\ldots,\ve'_{n_2}}$.
Here we place the spectral parameters and the spins of the
upper lines first, from left to right, and then the lower, also from
left to right. A clue to the difference equations is the following.
\proclaim Rotation.
\eqa
&&F^{(i)}(\zeta_1,\ldots,\zeta_{n_1};\zeta'_1,\ldots,\zeta'_{n_2})
_{\ve_1,\ldots,\ve_{n_1};\ve'_1,\ldots,\ve'_{n_2}} \nnb \\
&&~=
F^{(i+1)}(\zeta'_1/x,\zeta_1,\ldots,\zeta_{n_1};\zeta'_2,\ldots,\zeta'_{n_2})
_{-\ve'_1,\ve_1,\ldots,\ve_{n_1};\ve'_2,\ldots,\ve'_{n_2}},\label{eqn:rot1}\\
&&~=F^{(i)}(\zeta_1,\ldots,\zeta_{n_1-1};
\zeta'_1,\ldots,\zeta'_{n_2},\zeta_{n_1}/x)
_{\ve_1,\ldots,\ve_{n_1-1};\ve'_1,\ldots,\ve'_{n_2},-\ve_{n_1}}.
\label{eqn:rot2}
\endeqa
\par\noindent

Let us give heuristic arguments to derive \refeq{Diff}-\refeq{rot2}.
We suppose that the correlators are given as configuration sums
on the infinite lattice. It makes sense to do that because
of the normalization of the Boltzmann weights by $\kappa$.
We further assume that the vertices on the lattice that are pushed away
to infinity by a $Z$-invariant deformation of the lattice, can be neglected.
Let us use this assumption in a concrete situation.
Consider a part of the lattice as given in Figure 1.3, on the left.
The rest of the lattice, not shown in this figure, is irrelevant in the
following argument. Denote the configuration sum with the spins
$\ve_1$ and $\ve_2$ by $F(\ve_1,\ve_2;\Lambda)$. The configuration sum
for the lattice on the right, is given by
\eqa
&&F^{(i)}(\varepsilon_1,\varepsilon_2;\Lambda')
=\sum_{\varepsilon'_1,\varepsilon'_2}R(\zeta_1/\zeta_2)
_{\varepsilon_1,\varepsilon_2;\varepsilon'_1,\varepsilon'_2}
F^{(i)}(\varepsilon'_1,\varepsilon'_2;\Lambda).
\endeqa
\noindent
We can push out the extra vertex in $\Lambda'$ to the north as far as we wish
by using the Yang-Baxter equation, without changing the statistical sum
$F^{(i)}(\varepsilon_1,\varepsilon_2;\Lambda')$. Therefore by the above
assumption,
we have $F^{(i)}(\varepsilon_1,\varepsilon_2;\Lambda')=
F^{(i)}(\varepsilon_1,\varepsilon_2;\Lambda)$, showing \refeq{Symm}.

%\Fig(f13.tex)
\setlength{\unitlength}{0.0125in}
\begin{picture}(330,371)(0,-10)
\drawline(240,330)(240,160)(200,130)
\drawline(205.200,136.400)(200.000,130.000)(207.600,133.200)
\drawline(200,330)(200,160)(240,130)
\drawline(232.400,133.200)(240.000,130.000)(234.800,136.400)
\drawline(80,330)(80,130)
\drawline(78.000,138.000)(80.000,130.000)(82.000,138.000)
\drawline(40,330)(40,130)
\drawline(38.000,138.000)(40.000,130.000)(42.000,138.000)
\drawline(160,330)(160,90)(280,90)(280,330)
\drawline(0,330)(0,90)(120,90)(120,330)
\drawline(13.000,292.000)(5.000,290.000)(13.000,288.000)
\drawline(5,290)(115,290)
\drawline(13.000,252.000)(5.000,250.000)(13.000,248.000)
\drawline(5,250)(115,250)
\drawline(13.000,212.000)(5.000,210.000)(13.000,208.000)
\drawline(5,210)(115,210)
\drawline(13.000,172.000)(5.000,170.000)(13.000,168.000)
\drawline(5,170)(115,170)
\drawline(173.000,172.000)(165.000,170.000)(173.000,168.000)
\drawline(165,170)(275,170)
\drawline(173.000,212.000)(165.000,210.000)(173.000,208.000)
\drawline(165,210)(275,210)
\drawline(173.000,252.000)(165.000,250.000)(173.000,248.000)
\drawline(165,250)(275,250)
\drawline(173.000,292.000)(165.000,290.000)(173.000,288.000)
\drawline(165,290)(275,290)
\put(5,0){\makebox(0,0)[lb]{\raisebox{0pt}[0pt][0pt]{\shortstack[l]{{\twlrm up
to infinity.}}}}}
\put(5,15){\makebox(0,0)[lb]{\raisebox{0pt}[0pt][0pt]{\shortstack[l]{{\twlrm
The extra vertex in $\Lambda'$ can be pushed}}}}}
\put(5,30){\makebox(0,0)[lb]{\raisebox{0pt}[0pt][0pt]{\shortstack[l]{{\twlrm
Figure 1.3\quad Twisting two vertical lines.}}}}}
\put(80,115){\makebox(0,0)[lb]{\raisebox{0pt}[0pt][0pt]{\shortstack[l]{{\twlrm
$\varepsilon_2$}}}}}
\put(40,115){\makebox(0,0)[lb]{\raisebox{0pt}[0pt][0pt]{\shortstack[l]{{\twlrm
$\varepsilon_1$}}}}}
\put(80,345){\makebox(0,0)[lb]{\raisebox{0pt}[0pt][0pt]{\shortstack[l]{{\twlrm
$\zeta_1$}}}}}
\put(40,345){\makebox(0,0)[lb]{\raisebox{0pt}[0pt][0pt]{\shortstack[l]{{\twlrm
$\zeta_1$}}}}}
\put(200,115){\makebox(0,0)[lb]{\raisebox{0pt}[0pt][0pt]{\shortstack[l]{{\twlrm
$\varepsilon_2$}}}}}
\put(240,115){\makebox(0,0)[lb]{\raisebox{0pt}[0pt][0pt]{\shortstack[l]{{\twlrm
$\varepsilon_1$}}}}}
\put(220,65){\makebox(0,0)[lb]{\raisebox{0pt}[0pt][0pt]{\shortstack[l]{{\twlrm
$\Lambda'$}}}}}
\put(55,65){\makebox(0,0)[lb]{\raisebox{0pt}[0pt][0pt]{\shortstack[l]{{\twlrm
$\Lambda$}}}}}
\put(200,345){\makebox(0,0)[lb]{\raisebox{0pt}[0pt][0pt]{\shortstack[l]{{\twlrm
$\zeta_1$}}}}}
\put(240,345){\makebox(0,0)[lb]{\raisebox{0pt}[0pt][0pt]{\shortstack[l]{{\twlrm
$\zeta_2$}}}}}
\end{picture}

Next, see Figure 1.4. The right hand side of the figure represents the left
hand side of \refeq{Nor1}, the bullet $\bullet$ showing the summation
$\sum_{\ve=\pm}$. Using unitarity, as shown on the left hand side of
the figure, we can push away the two vertical lines with the spectral
parameters $\zeta$ and $x\zeta$. Therefore, we have \refeq{Nor1}.

Eq.\refeq{Nor2} follows from \refeq{Symm} and \refeq{Rx}.
Eq.\refeq{Z2i1} is obvious. Eq.\refeq{Z2i2} follows from the parity
of the Boltzmann weights.

Finally, we will show \refeq{rot1}-\refeq{rot2}.
Since the arguments are the same, we will consider only \refeq{rot1}.
See Figure 1.5, where $\zeta'_1$
in \refeq{rot1} is shown as $\zeta$. We can push away the semi circle
with the spectral parameter $\zeta$. In the limit, we have Eq.\refeq{rot1}.

%\Fig(f14.tex)
\setlength{\unitlength}{0.0125in}
\begin{picture}(393,360)(0,-10)
\drawline(355,115)(355,325)
\drawline(275,115)(275,325)
\drawline(268.000,287.000)(260.000,285.000)(268.000,283.000)
\drawline(260,285)(370,285)
\drawline(268.000,247.000)(260.000,245.000)(268.000,243.000)
\drawline(260,245)(370,245)
\drawline(268.000,207.000)(260.000,205.000)(268.000,203.000)
\drawline(260,205)(370,205)
\drawline(268.000,167.000)(260.000,165.000)(268.000,163.000)
\drawline(260,165)(370,165)
\drawline(268.000,127.000)(260.000,125.000)(268.000,123.000)
\drawline(260,125)(370,125)
\drawline(250,325)(250,80)(375,80)(375,325)
\drawline(63.000,287.000)(55.000,285.000)(63.000,283.000)
\drawline(55,285)(215,285)
\drawline(63.000,327.000)(55.000,325.000)(63.000,323.000)
\drawline(55,325)(215,325)
\drawline(355,115)	(352.193,112.740)
	(349.740,110.828)
	(345.721,107.926)
	(342.592,106.062)
	(340.000,105.000)

\drawline(340,105)	(337.716,104.539)
	(334.688,104.385)
	(330.567,104.539)
	(327.986,104.731)
	(325.000,105.000)

\drawline(333.154,106.226)(325.000,105.000)(332.772,102.245)
\drawline(275,115)	(277.807,112.740)
	(280.260,110.828)
	(284.279,107.926)
	(287.408,106.062)
	(290.000,105.000)

\drawline(290,105)	(292.284,104.539)
	(295.312,104.385)
	(299.433,104.539)
	(302.014,104.731)
	(305.000,105.000)

\drawline(297.228,102.245)(305.000,105.000)(296.846,106.226)
\drawline(116.805,84.081)(125.000,85.000)(117.337,88.045)
\drawline(125,85)	(122.370,85.358)
	(119.924,85.701)
	(115.552,86.350)
	(111.811,86.964)
	(108.627,87.556)
	(103.638,88.733)
	(100.000,90.000)

\drawline(100,90)	(96.082,91.841)
	(91.790,94.202)
	(86.979,97.183)
	(84.333,98.939)
	(81.502,100.889)
	(78.467,103.045)
	(75.212,105.421)
	(71.716,108.028)
	(67.962,110.881)
	(63.932,113.992)
	(59.607,117.374)
	(54.969,121.039)
	(50.000,125.000)

\drawline(152.663,88.045)(145.000,85.000)(153.195,84.081)
\drawline(145,85)	(147.630,85.358)
	(150.076,85.701)
	(154.448,86.350)
	(158.189,86.964)
	(161.373,87.556)
	(166.362,88.733)
	(170.000,90.000)

\drawline(170,90)	(173.918,91.841)
	(178.210,94.202)
	(183.021,97.183)
	(185.667,98.939)
	(188.498,100.889)
	(191.533,103.045)
	(194.788,105.421)
	(198.284,108.028)
	(202.038,110.881)
	(206.068,113.992)
	(210.393,117.374)
	(215.031,121.039)
	(220.000,125.000)

\drawline(60.262,91.349)(55.000,85.000)(62.631,88.126)
\drawline(55,85)	(58.767,87.754)
	(62.415,90.393)
	(65.949,92.918)
	(69.372,95.332)
	(72.687,97.636)
	(75.899,99.832)
	(79.010,101.921)
	(82.025,103.906)
	(84.948,105.789)
	(87.781,107.571)
	(90.530,109.254)
	(93.196,110.840)
	(95.784,112.331)
	(98.298,113.728)
	(103.117,116.250)
	(107.683,118.420)
	(112.024,120.254)
	(116.170,121.765)
	(120.150,122.969)
	(123.994,123.879)
	(127.731,124.512)
	(131.390,124.880)
	(135.000,125.000)

\drawline(135,125)	(138.610,124.880)
	(142.269,124.512)
	(146.006,123.879)
	(149.850,122.969)
	(153.830,121.765)
	(157.976,120.254)
	(162.317,118.420)
	(166.883,116.250)
	(171.702,113.728)
	(174.216,112.331)
	(176.804,110.840)
	(179.470,109.254)
	(182.219,107.571)
	(185.052,105.789)
	(187.975,103.906)
	(190.990,101.921)
	(194.101,99.832)
	(197.313,97.636)
	(200.628,95.332)
	(204.051,92.918)
	(207.585,90.393)
	(211.233,87.754)
	(215.000,85.000)

\drawline(62.631,221.874)(55.000,225.000)(60.262,218.651)
\drawline(55,225)	(58.767,222.246)
	(62.415,219.607)
	(65.949,217.082)
	(69.372,214.668)
	(72.687,212.364)
	(75.899,210.168)
	(79.010,208.079)
	(82.025,206.094)
	(84.948,204.211)
	(87.781,202.429)
	(90.530,200.746)
	(93.196,199.160)
	(95.784,197.669)
	(98.298,196.272)
	(103.117,193.750)
	(107.683,191.580)
	(112.024,189.746)
	(116.170,188.235)
	(120.150,187.031)
	(123.994,186.121)
	(127.731,185.488)
	(131.390,185.120)
	(135.000,185.000)

\drawline(135,185)	(138.610,185.120)
	(142.269,185.488)
	(146.006,186.121)
	(149.850,187.031)
	(153.830,188.235)
	(157.976,189.746)
	(162.317,191.580)
	(166.883,193.750)
	(171.702,196.272)
	(174.216,197.669)
	(176.804,199.160)
	(179.470,200.746)
	(182.219,202.429)
	(185.052,204.211)
	(187.975,206.094)
	(190.990,208.079)
	(194.101,210.168)
	(197.313,212.364)
	(200.628,214.668)
	(204.051,217.082)
	(207.585,219.607)
	(211.233,222.246)
	(215.000,225.000)

\drawline(60.262,191.349)(55.000,185.000)(62.631,188.126)
\drawline(55,185)	(58.767,187.754)
	(62.415,190.393)
	(65.949,192.918)
	(69.372,195.332)
	(72.687,197.636)
	(75.899,199.832)
	(79.010,201.921)
	(82.025,203.906)
	(84.948,205.789)
	(87.781,207.571)
	(90.530,209.254)
	(93.196,210.840)
	(95.784,212.331)
	(98.298,213.728)
	(103.117,216.250)
	(107.683,218.420)
	(112.024,220.254)
	(116.170,221.765)
	(120.150,222.969)
	(123.994,223.879)
	(127.731,224.512)
	(131.390,224.880)
	(135.000,225.000)

\drawline(135,225)	(138.610,224.880)
	(142.269,224.512)
	(146.006,223.879)
	(149.850,222.969)
	(153.830,221.765)
	(157.976,220.254)
	(162.317,218.420)
	(166.883,216.250)
	(171.702,213.728)
	(174.216,212.331)
	(176.804,210.840)
	(179.470,209.254)
	(182.219,207.571)
	(185.052,205.789)
	(187.975,203.906)
	(190.990,201.921)
	(194.101,199.832)
	(197.313,197.636)
	(200.628,195.332)
	(204.051,192.918)
	(207.585,190.393)
	(211.233,187.754)
	(215.000,185.000)

\put(15,0){\makebox(0,0)[lb]{\raisebox{0pt}[0pt][0pt]{\shortstack[l]{{\twlrm
left half of one of the lines are reversed.}}}}}
\put(15,15){\makebox(0,0)[lb]{\raisebox{0pt}[0pt][0pt]{\shortstack[l]{{\twlrm
oriented left, and the case where the orientation of the}}}}}
\put(15,30){\makebox(0,0)[lb]{\raisebox{0pt}[0pt][0pt]{\shortstack[l]{{\twlrm
Figure 1.4\quad The unitarity with the two horizontal lines}}}}}
\put(345,335){\makebox(0,0)[lb]{\raisebox{0pt}[0pt][0pt]{\shortstack[l]{{\twlrm
$x\zeta$}}}}}
\put(275,335){\makebox(0,0)[lb]{\raisebox{0pt}[0pt][0pt]{\shortstack[l]{{\twlrm
$\zeta$}}}}}
\put(315,105){\makebox(0,0)[lb]{\raisebox{0pt}[0pt][0pt]{\shortstack[l]{{\twlrm
$\bullet$}}}}}
\put(130,235){\makebox(0,0)[lb]{\raisebox{0pt}[0pt][0pt]{\shortstack[l]{{\twlrm
$\xi$}}}}}
\put(125,170){\makebox(0,0)[lb]{\raisebox{0pt}[0pt][0pt]{\shortstack[l]{{\twlrm
$x\zeta$}}}}}
\put(120,335){\makebox(0,0)[lb]{\raisebox{0pt}[0pt][0pt]{\shortstack[l]{{\twlrm
$x\zeta$}}}}}
\put(160,75){\makebox(0,0)[lb]{\raisebox{0pt}[0pt][0pt]{\shortstack[l]{{\twlrm
$x\zeta$}}}}}
\put(90,75){\makebox(0,0)[lb]{\raisebox{0pt}[0pt][0pt]{\shortstack[l]{{\twlrm
$\zeta$}}}}}
\put(130,295){\makebox(0,0)[lb]{\raisebox{0pt}[0pt][0pt]{\shortstack[l]{{\twlrm
$\xi$}}}}}
\put(135,85){\makebox(0,0)[lb]{\raisebox{0pt}[0pt][0pt]{\shortstack[l]{{\twlrm
$\bullet$}}}}}
\put(10,100){\makebox(0,0)[lb]{\raisebox{0pt}[0pt][0pt]{\shortstack[l]{{\twlrm
$=$}}}}}
\put(10,205){\makebox(0,0)[lb]{\raisebox{0pt}[0pt][0pt]{\shortstack[l]{{\twlrm
$=$}}}}}
\put(0,300){\makebox(0,0)[lb]{\raisebox{0pt}[0pt][0pt]{\shortstack[l]{{\twlrm
$id=$}}}}}
\end{picture}

%\Fig(f15.tex)
\setlength{\unitlength}{0.0125in}
\begin{picture}(365,530)(0,-10)
\drawline(240,515)(240,315)
\drawline(238.000,323.000)(240.000,315.000)(242.000,323.000)
\drawline(208.000,357.000)(200.000,355.000)(208.000,353.000)
\drawline(200,355)(320,355)
\drawline(208.000,477.000)(200.000,475.000)(208.000,473.000)
\drawline(200,475)(320,475)
\drawline(60,515)(60,315)
\drawline(58.000,323.000)(60.000,315.000)(62.000,323.000)
\drawline(28.000,357.000)(20.000,355.000)(28.000,353.000)
\drawline(20,355)(140,355)
\drawline(28.000,477.000)(20.000,475.000)(28.000,473.000)
\drawline(20,475)(140,475)
\drawline(100,435)(100,315)
\drawline(98.000,323.000)(100.000,315.000)(102.000,323.000)
\drawline(198.000,232.000)(190.000,230.000)(198.000,228.000)
\drawline(190,230)(310,230)
\drawline(198.000,112.000)(190.000,110.000)(198.000,108.000)
\drawline(190,110)(310,110)
\drawline(230,270)(230,70)
\drawline(228.000,78.000)(230.000,70.000)(232.000,78.000)
\drawline(60,270)(60,70)
\drawline(58.000,78.000)(60.000,70.000)(62.000,78.000)
\drawline(28.000,112.000)(20.000,110.000)(28.000,108.000)
\drawline(20,110)(140,110)
\drawline(28.000,232.000)(20.000,230.000)(28.000,228.000)
\drawline(20,230)(140,230)
\drawline(270,255)	(274.333,252.584)
	(278.025,250.410)
	(281.135,248.435)
	(283.721,246.615)
	(287.556,243.262)
	(290.000,240.000)

\drawline(290,240)	(291.570,236.809)
	(292.841,233.187)
	(293.827,229.000)
	(294.543,224.119)
	(294.804,221.376)
	(295.004,218.410)
	(295.143,215.204)
	(295.224,211.741)
	(295.248,208.006)
	(295.218,203.982)
	(295.134,199.652)
	(295.000,195.000)

\drawline(293.265,203.062)(295.000,195.000)(297.263,202.930)
\drawline(280,435)	(276.655,434.458)
	(273.432,433.912)
	(270.329,433.361)
	(267.343,432.804)
	(264.471,432.239)
	(261.710,431.666)
	(259.057,431.083)
	(256.511,430.489)
	(251.724,429.264)
	(247.328,427.982)
	(243.300,426.634)
	(239.619,425.210)
	(236.261,423.701)
	(233.207,422.098)
	(230.433,420.393)
	(227.917,418.575)
	(223.573,414.567)
	(220.000,410.000)

\drawline(220,410)	(217.986,406.603)
	(216.271,402.934)
	(214.846,399.031)
	(213.707,394.934)
	(212.847,390.683)
	(212.260,386.317)
	(211.939,381.874)
	(211.877,377.396)
	(212.070,372.921)
	(212.511,368.489)
	(213.192,364.138)
	(214.109,359.910)
	(215.254,355.842)
	(216.622,351.975)
	(218.206,348.348)
	(220.000,345.000)

\drawline(220,345)	(223.071,340.592)
	(226.906,336.548)
	(231.667,332.779)
	(234.445,330.971)
	(237.515,329.199)
	(240.897,327.452)
	(244.611,325.719)
	(248.678,323.989)
	(253.117,322.251)
	(257.948,320.494)
	(260.518,319.605)
	(263.193,318.707)
	(265.976,317.799)
	(268.870,316.880)
	(271.877,315.947)
	(275.000,315.000)

\drawline(266.763,315.383)(275.000,315.000)(267.912,319.215)
\drawline(250,255)	(246.291,254.619)
	(242.845,254.222)
	(239.648,253.806)
	(236.689,253.368)
	(233.953,252.903)
	(231.430,252.408)
	(226.965,251.313)
	(223.192,250.054)
	(220.009,248.602)
	(217.312,246.927)
	(215.000,245.000)

\drawline(215,245)	(211.563,240.774)
	(210.062,238.241)
	(208.696,235.484)
	(207.457,232.544)
	(206.338,229.462)
	(205.330,226.281)
	(204.426,223.041)
	(203.618,219.784)
	(202.899,216.552)
	(202.259,213.386)
	(201.693,210.327)
	(201.192,207.418)
	(200.748,204.700)
	(200.000,200.000)

\drawline(200,200)	(199.673,197.655)
	(199.397,194.996)
	(199.169,192.069)
	(198.989,188.918)
	(198.856,185.588)
	(198.766,182.125)
	(198.720,178.573)
	(198.716,174.976)
	(198.751,171.381)
	(198.826,167.831)
	(198.937,164.371)
	(199.084,161.047)
	(199.265,157.903)
	(199.480,154.984)
	(200.000,150.000)

\drawline(200,150)	(200.326,147.349)
	(200.654,144.359)
	(201.001,141.080)
	(201.384,137.562)
	(201.821,133.856)
	(202.328,130.012)
	(202.922,126.079)
	(203.621,122.109)
	(204.441,118.151)
	(205.399,114.255)
	(206.512,110.472)
	(207.797,106.851)
	(209.272,103.444)
	(210.952,100.299)
	(212.856,97.468)
	(215.000,95.000)

\drawline(215,95)	(218.650,91.838)
	(222.900,89.132)
	(225.300,87.936)
	(227.910,86.837)
	(230.750,85.831)
	(233.841,84.910)
	(237.202,84.071)
	(240.854,83.307)
	(244.817,82.613)
	(249.111,81.983)
	(253.756,81.413)
	(256.217,81.148)
	(258.773,80.896)
	(261.426,80.656)
	(264.180,80.427)
	(267.038,80.208)
	(270.000,80.000)

\drawline(261.883,78.543)(270.000,80.000)(262.153,82.534)
\drawline(125,195)	(125.231,197.608)
	(125.439,200.126)
	(125.787,204.900)
	(126.040,209.340)
	(126.199,213.464)
	(126.260,217.291)
	(126.223,220.839)
	(126.085,224.126)
	(125.844,227.171)
	(125.498,229.991)
	(125.046,232.606)
	(123.816,237.292)
	(122.139,241.375)
	(120.000,245.000)

\drawline(120,245)	(117.399,247.808)
	(113.511,250.254)
	(110.939,251.415)
	(107.868,252.573)
	(104.241,253.759)
	(100.000,255.000)

\drawline(108.243,254.769)(100.000,255.000)(107.164,250.917)
\drawline(80,255)	(76.291,254.619)
	(72.845,254.222)
	(69.648,253.806)
	(66.689,253.368)
	(63.953,252.903)
	(61.430,252.408)
	(56.965,251.313)
	(53.192,250.054)
	(50.009,248.602)
	(47.312,246.927)
	(45.000,245.000)

\drawline(45,245)	(41.563,240.774)
	(40.062,238.241)
	(38.696,235.484)
	(37.457,232.544)
	(36.338,229.462)
	(35.330,226.281)
	(34.426,223.041)
	(33.618,219.784)
	(32.899,216.552)
	(32.259,213.386)
	(31.693,210.327)
	(31.192,207.418)
	(30.748,204.700)
	(30.000,200.000)

\drawline(30,200)	(29.673,197.655)
	(29.397,194.996)
	(29.169,192.069)
	(28.989,188.918)
	(28.856,185.588)
	(28.766,182.125)
	(28.720,178.573)
	(28.716,174.976)
	(28.751,171.381)
	(28.826,167.831)
	(28.937,164.371)
	(29.084,161.047)
	(29.265,157.903)
	(29.480,154.984)
	(30.000,150.000)

\drawline(30,150)	(30.326,147.349)
	(30.654,144.359)
	(31.001,141.080)
	(31.384,137.562)
	(31.821,133.856)
	(32.328,130.012)
	(32.922,126.079)
	(33.621,122.109)
	(34.441,118.151)
	(35.399,114.255)
	(36.512,110.472)
	(37.797,106.851)
	(39.272,103.444)
	(40.952,100.299)
	(42.856,97.468)
	(45.000,95.000)

\drawline(45,95)	(48.650,91.838)
	(52.900,89.132)
	(55.300,87.936)
	(57.910,86.837)
	(60.750,85.831)
	(63.840,84.910)
	(67.202,84.071)
	(70.854,83.307)
	(74.817,82.613)
	(79.111,81.983)
	(83.756,81.413)
	(86.217,81.148)
	(88.773,80.896)
	(91.426,80.656)
	(94.180,80.427)
	(97.038,80.208)
	(100.000,80.000)

\drawline(91.883,78.543)(100.000,80.000)(92.153,82.534)
\put(35,0){\makebox(0,0)[lb]{\raisebox{0pt}[0pt][0pt]{\shortstack[l]{{\twlrm
explicitly. Notice that after rotating the GSC changes.}}}}}
\put(185,100){\makebox(0,0)[lb]{\raisebox{0pt}[0pt][0pt]{\shortstack[l]{{\twlrm
$=$}}}}}
\put(10,100){\makebox(0,0)[lb]{\raisebox{0pt}[0pt][0pt]{\shortstack[l]{{\twlrm
$=$}}}}}
\put(255,220){\makebox(0,0)[lb]{\raisebox{0pt}[0pt][0pt]{\shortstack[l]{{\twlrm
$+$}}}}}
\put(275,245){\makebox(0,0)[lb]{\raisebox{0pt}[0pt][0pt]{\shortstack[l]{{\twlrm
$+$}}}}}
\put(190,175){\makebox(0,0)[lb]{\raisebox{0pt}[0pt][0pt]{\shortstack[l]{{\twlrm
$-$}}}}}
\put(215,220){\makebox(0,0)[lb]{\raisebox{0pt}[0pt][0pt]{\shortstack[l]{{\twlrm
$-$}}}}}
\put(235,235){\makebox(0,0)[lb]{\raisebox{0pt}[0pt][0pt]{\shortstack[l]{{\twlrm
$-$}}}}}
\put(240,245){\makebox(0,0)[lb]{\raisebox{0pt}[0pt][0pt]{\shortstack[l]{{\twlrm
$-$}}}}}
\put(320,230){\makebox(0,0)[lb]{\raisebox{0pt}[0pt][0pt]{\shortstack[l]{{\twlrm
$-$}}}}}
\put(295,185){\makebox(0,0)[lb]{\raisebox{0pt}[0pt][0pt]{\shortstack[l]{{\twlrm
$-$}}}}}
\put(275,75){\makebox(0,0)[lb]{\raisebox{0pt}[0pt][0pt]{\shortstack[l]{{\twlrm
$-$}}}}}
\put(240,100){\makebox(0,0)[lb]{\raisebox{0pt}[0pt][0pt]{\shortstack[l]{{\twlrm
$-$}}}}}
\put(285,100){\makebox(0,0)[lb]{\raisebox{0pt}[0pt][0pt]{\shortstack[l]{{\twlrm
$-$}}}}}
\put(230,60){\makebox(0,0)[lb]{\raisebox{0pt}[0pt][0pt]{\shortstack[l]{{\twlrm
$+$}}}}}
\put(210,85){\makebox(0,0)[lb]{\raisebox{0pt}[0pt][0pt]{\shortstack[l]{{\twlrm
$+$}}}}}
\put(220,115){\makebox(0,0)[lb]{\raisebox{0pt}[0pt][0pt]{\shortstack[l]{{\twlrm
$+$}}}}}
\put(240,165){\makebox(0,0)[lb]{\raisebox{0pt}[0pt][0pt]{\shortstack[l]{{\twlrm
$+$}}}}}
\put(180,225){\makebox(0,0)[lb]{\raisebox{0pt}[0pt][0pt]{\shortstack[l]{{\twlrm
$+$}}}}}
\put(210,250){\makebox(0,0)[lb]{\raisebox{0pt}[0pt][0pt]{\shortstack[l]{{\twlrm
$+$}}}}}
\put(230,280){\makebox(0,0)[lb]{\raisebox{0pt}[0pt][0pt]{\shortstack[l]{{\twlrm
$+$}}}}}
\put(120,185){\makebox(0,0)[lb]{\raisebox{0pt}[0pt][0pt]{\shortstack[l]{{\twlrm
$+$}}}}}
\put(20,175){\makebox(0,0)[lb]{\raisebox{0pt}[0pt][0pt]{\shortstack[l]{{\twlrm
$-$}}}}}
\put(45,220){\makebox(0,0)[lb]{\raisebox{0pt}[0pt][0pt]{\shortstack[l]{{\twlrm
$-$}}}}}
\put(65,240){\makebox(0,0)[lb]{\raisebox{0pt}[0pt][0pt]{\shortstack[l]{{\twlrm
$-$}}}}}
\put(85,245){\makebox(0,0)[lb]{\raisebox{0pt}[0pt][0pt]{\shortstack[l]{{\twlrm
$-$}}}}}
\put(150,230){\makebox(0,0)[lb]{\raisebox{0pt}[0pt][0pt]{\shortstack[l]{{\twlrm
$-$}}}}}
\put(100,70){\makebox(0,0)[lb]{\raisebox{0pt}[0pt][0pt]{\shortstack[l]{{\twlrm
$-$}}}}}
\put(65,95){\makebox(0,0)[lb]{\raisebox{0pt}[0pt][0pt]{\shortstack[l]{{\twlrm
$-$}}}}}
\put(110,100){\makebox(0,0)[lb]{\raisebox{0pt}[0pt][0pt]{\shortstack[l]{{\twlrm
$-$}}}}}
\put(55,60){\makebox(0,0)[lb]{\raisebox{0pt}[0pt][0pt]{\shortstack[l]{{\twlrm
$+$}}}}}
\put(35,90){\makebox(0,0)[lb]{\raisebox{0pt}[0pt][0pt]{\shortstack[l]{{\twlrm
$+$}}}}}
\put(45,115){\makebox(0,0)[lb]{\raisebox{0pt}[0pt][0pt]{\shortstack[l]{{\twlrm
$+$}}}}}
\put(90,220){\makebox(0,0)[lb]{\raisebox{0pt}[0pt][0pt]{\shortstack[l]{{\twlrm
$+$}}}}}
\put(65,165){\makebox(0,0)[lb]{\raisebox{0pt}[0pt][0pt]{\shortstack[l]{{\twlrm
$+$}}}}}
\put(10,230){\makebox(0,0)[lb]{\raisebox{0pt}[0pt][0pt]{\shortstack[l]{{\twlrm
$+$}}}}}
\put(40,250){\makebox(0,0)[lb]{\raisebox{0pt}[0pt][0pt]{\shortstack[l]{{\twlrm
$+$}}}}}
\put(60,280){\makebox(0,0)[lb]{\raisebox{0pt}[0pt][0pt]{\shortstack[l]{{\twlrm
$+$}}}}}
\put(235,305){\makebox(0,0)[lb]{\raisebox{0pt}[0pt][0pt]{\shortstack[l]{{\twlrm
$+$}}}}}
\put(215,330){\makebox(0,0)[lb]{\raisebox{0pt}[0pt][0pt]{\shortstack[l]{{\twlrm
$+$}}}}}
\put(225,360){\makebox(0,0)[lb]{\raisebox{0pt}[0pt][0pt]{\shortstack[l]{{\twlrm
$+$}}}}}
\put(245,385){\makebox(0,0)[lb]{\raisebox{0pt}[0pt][0pt]{\shortstack[l]{{\twlrm
$+$}}}}}
\put(285,430){\makebox(0,0)[lb]{\raisebox{0pt}[0pt][0pt]{\shortstack[l]{{\twlrm
$+$}}}}}
\put(270,300){\makebox(0,0)[lb]{\raisebox{0pt}[0pt][0pt]{\shortstack[l]{{\twlrm
$-$}}}}}
\put(245,340){\makebox(0,0)[lb]{\raisebox{0pt}[0pt][0pt]{\shortstack[l]{{\twlrm
$-$}}}}}
\put(280,345){\makebox(0,0)[lb]{\raisebox{0pt}[0pt][0pt]{\shortstack[l]{{\twlrm
$-$}}}}}
\put(195,340){\makebox(0,0)[lb]{\raisebox{0pt}[0pt][0pt]{\shortstack[l]{{\twlrm
$-$}}}}}
\put(205,405){\makebox(0,0)[lb]{\raisebox{0pt}[0pt][0pt]{\shortstack[l]{{\twlrm
$-$}}}}}
\put(230,445){\makebox(0,0)[lb]{\raisebox{0pt}[0pt][0pt]{\shortstack[l]{{\twlrm
$-$}}}}}
\put(280,465){\makebox(0,0)[lb]{\raisebox{0pt}[0pt][0pt]{\shortstack[l]{{\twlrm
$-$}}}}}
\put(210,465){\makebox(0,0)[lb]{\raisebox{0pt}[0pt][0pt]{\shortstack[l]{{\twlrm
$+$}}}}}
\put(230,500){\makebox(0,0)[lb]{\raisebox{0pt}[0pt][0pt]{\shortstack[l]{{\twlrm
$+$}}}}}
\put(45,320){\makebox(0,0)[lb]{\raisebox{0pt}[0pt][0pt]{\shortstack[l]{{\twlrm
$+$}}}}}
\put(75,345){\makebox(0,0)[lb]{\raisebox{0pt}[0pt][0pt]{\shortstack[l]{{\twlrm
$+$}}}}}
\put(90,390){\makebox(0,0)[lb]{\raisebox{0pt}[0pt][0pt]{\shortstack[l]{{\twlrm
$+$}}}}}
\put(25,465){\makebox(0,0)[lb]{\raisebox{0pt}[0pt][0pt]{\shortstack[l]{{\twlrm
$+$}}}}}
\put(50,500){\makebox(0,0)[lb]{\raisebox{0pt}[0pt][0pt]{\shortstack[l]{{\twlrm
$+$}}}}}
\put(125,345){\makebox(0,0)[lb]{\raisebox{0pt}[0pt][0pt]{\shortstack[l]{{\twlrm
$-$}}}}}
\put(85,320){\makebox(0,0)[lb]{\raisebox{0pt}[0pt][0pt]{\shortstack[l]{{\twlrm
$-$}}}}}
\put(20,340){\makebox(0,0)[lb]{\raisebox{0pt}[0pt][0pt]{\shortstack[l]{{\twlrm
$-$}}}}}
\put(100,465){\makebox(0,0)[lb]{\raisebox{0pt}[0pt][0pt]{\shortstack[l]{{\twlrm
$-$}}}}}
\put(50,415){\makebox(0,0)[lb]{\raisebox{0pt}[0pt][0pt]{\shortstack[l]{{\twlrm
$-$}}}}}
\put(35,15){\makebox(0,0)[lb]{\raisebox{0pt}[0pt][0pt]{\shortstack[l]{{\twlrm
to rotate a half line. A choice of GSC is written}}}}}
\put(35,30){\makebox(0,0)[lb]{\raisebox{0pt}[0pt][0pt]{\shortstack[l]{{\twlrm
Figure 1.5\quad The use of the Yang-Baxter equation}}}}}
\put(190,380){\makebox(0,0)[lb]{\raisebox{0pt}[0pt][0pt]{\shortstack[l]{{\twlrm
$\zeta$}}}}}
\put(110,195){\makebox(0,0)[lb]{\raisebox{0pt}[0pt][0pt]{\shortstack[l]{{\twlrm
$\zeta$}}}}}
\put(35,175){\makebox(0,0)[lb]{\raisebox{0pt}[0pt][0pt]{\shortstack[l]{{\twlrm
$\zeta$}}}}}
\put(110,430){\makebox(0,0)[lb]{\raisebox{0pt}[0pt][0pt]{\shortstack[l]{{\twlrm
$\zeta$}}}}}
\put(270,195){\makebox(0,0)[lb]{\raisebox{0pt}[0pt][0pt]{\shortstack[l]{{\twlrm
$\zeta/x$}}}}}
\put(210,175){\makebox(0,0)[lb]{\raisebox{0pt}[0pt][0pt]{\shortstack[l]{{\twlrm
$\zeta$}}}}}
\put(90,255){\makebox(0,0)[lb]{\raisebox{0pt}[0pt][0pt]{\shortstack[l]{{\twlrm
$\bullet$}}}}}
\put(260,255){\makebox(0,0)[lb]{\raisebox{0pt}[0pt][0pt]{\shortstack[l]{{\twlrm
$\bullet$}}}}}
\put(165,405){\makebox(0,0)[lb]{\raisebox{0pt}[0pt][0pt]{\shortstack[l]{{\twlrm
$=$}}}}}
\put(170,160){\makebox(0,0)[lb]{\raisebox{0pt}[0pt][0pt]{\shortstack[l]{{\twlrm
$=$}}}}}
\put(0,160){\makebox(0,0)[lb]{\raisebox{0pt}[0pt][0pt]{\shortstack[l]{{\twlrm
$=$}}}}}
\end{picture}

\def\ze{\zeta}
\def\ep{\varepsilon}

\section{Spontaneous staggered polarization}

Let us work out the case $n=2$ in detail.
Since $F^{(i)}_2(\ze_1,\ze_2)$ depends only on the ratio $\ze=\ze_1/\ze_2$,
we shall denote $F^{(i)}_2(\ze_1,\ze_2)$ by $F^{(i)}_2(\ze)$.
Set
$G^\pm(\ze)=F^{(0)}_2(\ze)\pm F^{(1)}_2(\ze)$.
Eq.\refeq{Diff} for $F^{(i)}_2(\ze)$ reads
$G^\pm(x^{-2}\ze)=\pm R(\ze)G^\pm(\ze)$.
Using \refeq{Z2i1}, \refeq{Z2i2} we have
$G^\pm_{+-}(\ze)=\pm G^\pm_{-+}(\ze)$,
$G^\pm_{+-}(-\ze)=G^\mp_{+-}(\ze)$.
Thus the equations reduce to
\eqa
&&G^+_{+-}(x^{-2}\ze)=\kab(\ze^2)^{-1}
{(x\ze^{-1};q)_\infty(qx^{-1}\ze;q)_\infty\over
(x\ze;q)_\infty(qx^{-1}\ze^{-1};q)_\infty}
G^+_{+-}(\ze).
\label{eqn:5}
\endeqa
Set
\eq
&&\varphi(z)=g(z)g(x^{-4}z^{-1}),\qquad
g(z)=
{(x^6z;q,x^4,x^4)_\infty(qx^6z;q,x^4,x^4)_\infty
\over
(x^4z^{-1};q,x^4,x^4)_\infty(qz^{-1};q,x^4,x^4)_\infty}
\endeq
which solves the equation
$
\varphi(x^{-4}z)=\kab(z)^{-1}\varphi(z).
$
It is easy to see that the following is a solution of \refeq{5}
satisfying the normalization \refeq{Nor1}, i.e.,
$G^+_{+-}(x^{-1})=1$:
\eqa
&&G^+_{+-}(\ze)=
{\varphi(\ze^2) \over \varphi(x^{-2})}
{(x^2;x^2)^2_\infty \over (q;q)^2_\infty}
{(qx\ze;q)_\infty(qx^{-1}\ze^{-1};q)_\infty
\over
(x^3\ze;x^2)_\infty(x\ze^{-1};x^2)_\infty}.
\label{eqn:G}
\endeqa
Moreover the solution is unique if we impose further the condition that
$\sqrt{q}<x$ and
\eq
&&G^+_{+-}(\ze) \hbox{ is holomorphic in the neighborhood of
$x\le |\ze| \le x^{-1}$}.
\endeq
In view of \refeq{correl}
and following the definition given in \cite{Bax},
the staggared polarization $P_0$ is
\eq
&&P_0={G^-_{+-}(x^{-1})\over G^+_{+-}(x^{-1})}=
{(-q;q)^2_\infty(x^2;x^2)^2_\infty
\over
(q;q)^2_\infty(-x^2;x^2)^2_\infty}.
\endeq
This formula was conjectured by Baxter and Kelland \cite{BK}.
It was proved in the trigonometric case $q=0$ by Baxter \cite{Bax}.
Our result is stronger than this: In the context of this section,
it says that the ratio of the configuration sums
for the lattice $\Lambda_a$ in Figure 2.1, in the $i=0$ sector,
with the two different choices of the spins at the end edges, i.e.,
$(\varepsilon_1,\varepsilon_2)=(\pm,\mp)$,
is given by
\eq
&&{1\over2}{G^+_{+-}(\zeta)+G^-_{+-}(\zeta)
\over G^+_{-+}(\zeta)-G^-_{-+}(\zeta)},
\endeq
and that of the lattice $\Lambda_b$ is given by
\eq
&&{1\over2}{G^+_{+-}(\zeta/x)+G^-_{+-}(\zeta/x)
\over G^+_{-+}(\zeta/x)-G^-_{-+}(\zeta/x)}.
\endeq
We have checked these statements to a first few orders in the low temperature
expansions.

%\Fig(f21.tex)
\setlength{\unitlength}{0.0125in}
\begin{picture}(360,226)(0,-10)
\drawline(280,110)(280,60)
\drawline(278.000,68.000)(280.000,60.000)(282.000,68.000)
\drawline(280,180)(280,130)
\drawline(278.000,138.000)(280.000,130.000)(282.000,138.000)
\drawline(340,180)(340,60)
\drawline(338.000,68.000)(340.000,60.000)(342.000,68.000)
\drawline(220,180)(220,60)
\drawline(218.000,68.000)(220.000,60.000)(222.000,68.000)
\drawline(208.000,82.000)(200.000,80.000)(208.000,78.000)
\drawline(200,80)(360,80)
\drawline(208.000,162.000)(200.000,160.000)(208.000,158.000)
\drawline(200,160)(360,160)
\drawline(100,180)(100,120)
\drawline(98.000,128.000)(100.000,120.000)(102.000,128.000)
\drawline(60,180)(60,120)
\drawline(58.000,128.000)(60.000,120.000)(62.000,128.000)
\drawline(140,180)(140,60)
\drawline(138.000,68.000)(140.000,60.000)(142.000,68.000)
\drawline(20,180)(20,60)
\drawline(18.000,68.000)(20.000,60.000)(22.000,68.000)
\drawline(8.000,82.000)(0.000,80.000)(8.000,78.000)
\drawline(0,80)(160,80)
\drawline(8.000,162.000)(0.000,160.000)(8.000,158.000)
\drawline(0,160)(160,160)
\put(5,0){\makebox(0,0)[lb]{\raisebox{0pt}[0pt][0pt]{\shortstack[l]{{\twlrm the
$i=0$ GSC.}}}}}
\put(210,120){\makebox(0,0)[lb]{\raisebox{0pt}[0pt][0pt]{\shortstack[l]{{\twlrm
$-$}}}}}
\put(10,120){\makebox(0,0)[lb]{\raisebox{0pt}[0pt][0pt]{\shortstack[l]{{\twlrm
$-$}}}}}
\put(270,130){\makebox(0,0)[lb]{\raisebox{0pt}[0pt][0pt]{\shortstack[l]{{\twlrm
$\varepsilon_2$}}}}}
\put(270,105){\makebox(0,0)[lb]{\raisebox{0pt}[0pt][0pt]{\shortstack[l]{{\twlrm
$\varepsilon_1$}}}}}
\put(100,110){\makebox(0,0)[lb]{\raisebox{0pt}[0pt][0pt]{\shortstack[l]{{\twlrm
$\varepsilon_2$}}}}}
\put(60,110){\makebox(0,0)[lb]{\raisebox{0pt}[0pt][0pt]{\shortstack[l]{{\twlrm
$\varepsilon_1$}}}}}
\put(220,200){\makebox(0,0)[lb]{\raisebox{0pt}[0pt][0pt]{\shortstack[l]{{\twlrm
$\Lambda_b$}}}}}
\put(20,200){\makebox(0,0)[lb]{\raisebox{0pt}[0pt][0pt]{\shortstack[l]{{\twlrm
$\Lambda_a$}}}}}
\put(5,15){\makebox(0,0)[lb]{\raisebox{0pt}[0pt][0pt]{\shortstack[l]{{\twlrm
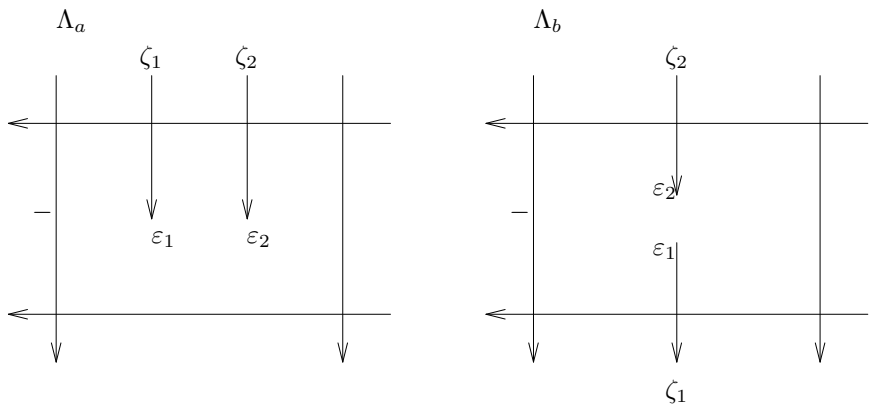
Figure 2.1\quad Two types of dislocation. The $-$ shows}}}}}
\put(95,185){\makebox(0,0)[lb]{\raisebox{0pt}[0pt][0pt]{\shortstack[l]{{\twlrm
$\zeta_2$}}}}}
\put(55,185){\makebox(0,0)[lb]{\raisebox{0pt}[0pt][0pt]{\shortstack[l]{{\twlrm
$\zeta_1$}}}}}
\put(275,185){\makebox(0,0)[lb]{\raisebox{0pt}[0pt][0pt]{\shortstack[l]{{\twlrm
$\zeta_2$}}}}}
\put(275,45){\makebox(0,0)[lb]{\raisebox{0pt}[0pt][0pt]{\shortstack[l]{{\twlrm
$\zeta_1$}}}}}
\end{picture}

%\input sec3
%		Difference equations for correlation functions
%		of the eight vertex model
%

\def\H{{\cal H}}

\section{Vertex operators}

In this section we shall reformulate the construction of
\sec(1) as an operator theory.
To this end let us first recall the corner transfer matrices (CTMs) following
\cite{Baxbk}.
Consider the lattice drawn in Figure 3.1.
Number the rows (resp. columns) from bottom-to-top (resp. from right-to-left)
as $-N+1, \cdots, 0, 1, \cdots, N$.
Fixing the boundary spins to the $i$-th GSC,
we denote by $A_0(\zt)$, $A_1(\zt)$, $A_2(\zt)$, $A_3(\zt)$ the CTM
corresponding to the NE, SE, SW, NW quadrant respectively.
When formulated in the IRF language
they are the transpose of $B$, $A$, $D$, $C$ in \cite{Baxbk}, p. 366.
For instance the $(\sigma,\sigma')$ element of $A_3(\zt)$ is the partition
function of the NW quadrant where the horizontal (resp. vertical) boundary
spins are fixed to $\sigma=(\vep_1,\cdots,\vep_N)$
(resp. $\sigma'=(\vep'_1,\cdots,\vep'_N)$). From the relations
\refeq{cross} it follows that
$A_2(\zt)={\cal R}A_3(x^{-1}\zt^{-1})$, $A_1(\zt)={\cal R}A_3(\zt){\cal R}$,
$A_0(\zt)=A_3(x^{-1}\zt^{-1}){\cal R}$,
where ${\cal R}=\sigma^x\otimes\cdots\otimes\sigma^x$ denotes
the spin reversal operator.
Normalize the CTMs so that the largest eigenvalue of $A_3(\zt)$ is $1$.
Baxter's argument shows that in the principal regime we have
$\lim_{N\rightarrow \infty} A_3(\zt)=\zt^{-D}$,
where the operator $D$ is independent of $\zt$ and has discrete eigenvalues
$0, 1, 2, \cdots$.
Consequently
\eqa
&&\lim_{N\rightarrow \infty} A_0(\zt)A_1(\zt)A_2(\zt)A_3(\zt)=x^{2D}.
\label{eqn:CTM}
\endeqa

%\Fig(f31.tex)
\setlength{\unitlength}{0.0125in}
\begin{picture}(350,379)(0,-10)
\drawline(258.000,148.000)(260.000,140.000)(262.000,148.000)
\drawline(260,140)(260,265)
\drawline(218.000,108.000)(220.000,100.000)(222.000,108.000)
\drawline(220,100)(220,305)
\drawline(215,100)(215,100)
\drawline(58.000,148.000)(60.000,140.000)(62.000,148.000)
\drawline(60,140)(60,260)
\drawline(98.000,108.000)(100.000,100.000)(102.000,108.000)
\drawline(100,100)(100,300)
\drawline(108.000,302.000)(100.000,300.000)(108.000,298.000)
\drawline(100,300)(220,300)
\drawline(68.000,262.000)(60.000,260.000)(68.000,258.000)
\drawline(60,260)(260,260)
\drawline(28.000,222.000)(20.000,220.000)(28.000,218.000)
\drawline(20,220)(300,220)
\drawline(28.000,182.000)(20.000,180.000)(28.000,178.000)
\drawline(20,180)(300,180)
\drawline(68.000,142.000)(60.000,140.000)(68.000,138.000)
\drawline(60,140)(260,140)
\drawline(108.000,102.000)(100.000,100.000)(108.000,98.000)
\drawline(100,100)(220,100)
\drawline(95,105)(95,105)
\drawline(178.000,73.000)(180.000,65.000)(182.000,73.000)
\drawline(180,65)(180,345)
\drawline(138.000,68.000)(140.000,60.000)(142.000,68.000)
\drawline(140,60)(140,345)
\put(20,0){\makebox(0,0)[lb]{\raisebox{0pt}[0pt][0pt]{\shortstack[l]{{\tenrm
The $0$-th GSC is depicted.}}}}}
\put(20,15){\makebox(0,0)[lb]{\raisebox{0pt}[0pt][0pt]
{\shortstack[l]{{\tenrm {Figure 3.1}\quad The CTMs $A_0,\cdots,A_3$.}}}}}
\put(20,35){\makebox(0,0)[lb]{\raisebox{0pt}[0pt][0pt]{\shortstack[l]{{\twlrm
$A_2$}}}}}
\put(320,55){\makebox(0,0)[lb]{\raisebox{0pt}[0pt][0pt]{\shortstack[l]{{\twlrm
$A_1$}}}}}
\put(320,330){\makebox(0,0)[lb]{\raisebox{0pt}[0pt][0pt]{\shortstack[l]{{\twlrm
$A_0$}}}}}
\put(0,335){\makebox(0,0)[lb]{\raisebox{0pt}[0pt][0pt]{\shortstack[l]{{\twlrm
$A_3$}}}}}
\put(110,305){\makebox(0,0)[lb]{\raisebox{0pt}[0pt][0pt]{\shortstack[l]{{\twlrm
$+$}}}}}
\put(90,280){\makebox(0,0)[lb]{\raisebox{0pt}[0pt][0pt]{\shortstack[l]{{\twlrm
$+$}}}}}
\put(45,235){\makebox(0,0)[lb]{\raisebox{0pt}[0pt][0pt]{\shortstack[l]{{\twlrm
$+$}}}}}
\put(70,260){\makebox(0,0)[lb]{\raisebox{0pt}[0pt][0pt]{\shortstack[l]{{\twlrm
$+$}}}}}
\put(0,215){\makebox(0,0)[lb]{\raisebox{0pt}[0pt][0pt]{\shortstack[l]{{\twlrm
$+$}}}}}
\put(0,175){\makebox(0,0)[lb]{\raisebox{0pt}[0pt][0pt]{\shortstack[l]{{\twlrm
$-$}}}}}
\put(45,155){\makebox(0,0)[lb]{\raisebox{0pt}[0pt][0pt]{\shortstack[l]{{\twlrm
$+$}}}}}
\put(70,125){\makebox(0,0)[lb]{\raisebox{0pt}[0pt][0pt]{\shortstack[l]{{\twlrm
$-$}}}}}
\put(90,115){\makebox(0,0)[lb]{\raisebox{0pt}[0pt][0pt]{\shortstack[l]{{\twlrm
$+$}}}}}
\put(115,85){\makebox(0,0)[lb]{\raisebox{0pt}[0pt][0pt]{\shortstack[l]{{\twlrm
$-$}}}}}
\put(135,45){\makebox(0,0)[lb]{\raisebox{0pt}[0pt][0pt]{\shortstack[l]{{\twlrm
$+$}}}}}
\put(175,50){\makebox(0,0)[lb]{\raisebox{0pt}[0pt][0pt]{\shortstack[l]{{\twlrm
$-$}}}}}
\put(225,110){\makebox(0,0)[lb]{\raisebox{0pt}[0pt][0pt]{\shortstack[l]{{\twlrm
$-$}}}}}
\put(195,85){\makebox(0,0)[lb]{\raisebox{0pt}[0pt][0pt]{\shortstack[l]{{\twlrm
$-$}}}}}
\put(230,125){\makebox(0,0)[lb]{\raisebox{0pt}[0pt][0pt]{\shortstack[l]{{\twlrm
$-$}}}}}
\put(265,155){\makebox(0,0)[lb]{\raisebox{0pt}[0pt][0pt]{\shortstack[l]{{\twlrm
$-$}}}}}
\put(310,175){\makebox(0,0)[lb]{\raisebox{0pt}[0pt][0pt]{\shortstack[l]{{\twlrm
$-$}}}}}
\put(310,215){\makebox(0,0)[lb]{\raisebox{0pt}[0pt][0pt]{\shortstack[l]{{\twlrm
$+$}}}}}
\put(265,235){\makebox(0,0)[lb]{\raisebox{0pt}[0pt][0pt]{\shortstack[l]{{\twlrm
$-$}}}}}
\put(235,260){\makebox(0,0)[lb]{\raisebox{0pt}[0pt][0pt]{\shortstack[l]{{\twlrm
$+$}}}}}
\put(225,275){\makebox(0,0)[lb]{\raisebox{0pt}[0pt][0pt]{\shortstack[l]{{\twlrm
$-$}}}}}
\put(195,300){\makebox(0,0)[lb]{\raisebox{0pt}[0pt][0pt]{\shortstack[l]{{\twlrm
$+$}}}}}
\put(175,355){\makebox(0,0)[lb]{\raisebox{0pt}[0pt][0pt]{\shortstack[l]{{\twlrm
$-$}}}}}
\put(130,355){\makebox(0,0)[lb]{\raisebox{0pt}[0pt][0pt]{\shortstack[l]{{\twlrm
$+$}}}}}
\end{picture}

Denote by $\H_i$ the space of eigenvectors of $A_3(\zt)$ in the limit.
In the trigonometric case $q=0$, as was shown in \cite{FM,DFJMN},
$\H_i$ can be identified
with the level-1 highest weight module $V(\La_i)$ of the quantized
affine algebra $U_{-x}(\slth)$ (with $-x$ playing the role of the deformation
parameter), and $D$ acts as the grading operator in the prinicipal gradation.
In the elliptic case, we have no such representation theoretical picture.
However if $\H_i=\oplus_{d=0}^{\infty} \H_{i,d}$ is the
eigenspace decomposition for $D$, then $\dim \H_{i,d}$ cannot change
continuously. Hence it should be the same as the $q=0$ case, giving
\eqa
&&\tr_{\H_i}x^{2D}=\sum_{d=0}^\infty x^{2d}\dim \H_{i,d}
= {1\over (x^2;x^4)_\infty}. \label{eqn:character}
\endeqa

%\Fig(f32.tex)
\setlength{\unitlength}{0.0125in}
\begin{picture}(588,415)(0,-10)
\drawline(68.000,127.000)(60.000,125.000)(68.000,123.000)
\drawline(60,125)(145,125)
\drawline(68.000,287.000)(60.000,285.000)(68.000,283.000)
\drawline(60,285)(140,285)
\drawline(68.000,327.000)(60.000,325.000)(68.000,323.000)
\drawline(60,325)(145,325)
\drawline(68.000,167.000)(60.000,165.000)(68.000,163.000)
\drawline(60,165)(140,165)
\drawline(98.000,93.000)(100.000,85.000)(102.000,93.000)
\drawline(100,85)(100,370)
\put(0,0){\makebox(0,0)[lb]{\raisebox{0pt}[0pt][0pt]{\shortstack[l]{{\twlrm
where $\sigma=(\varepsilon_1,\cdots,\varepsilon_n)$,
$\sigma'=(\varepsilon'_1,\cdots,\varepsilon'_n)$.}}}}}
\put(0,15){\makebox(0,0)[lb]{\raisebox{0pt}[0pt][0pt]
{\shortstack[l]{{\twlrm {Figure 3.2}
\quad Matrix element $\bigl(\Phi^{i+1}_{i\,\varepsilon}
(\zeta)\bigr)_{\sigma\sigma'}$}}}}}
\put(40,130){\makebox(0,0)[lb]{\raisebox{0pt}[0pt][0pt]{\shortstack[l]{{\twlrm
$\varepsilon_1$}}}}}
\put(35,175){\makebox(0,0)[lb]{\raisebox{0pt}[0pt][0pt]{\shortstack[l]{{\twlrm
$\varepsilon_2$}}}}}
\put(30,335){\makebox(0,0)[lb]{\raisebox{0pt}[0pt][0pt]{\shortstack[l]{{\twlrm
$\varepsilon_n$}}}}}
\put(160,320){\makebox(0,0)[lb]{\raisebox{0pt}[0pt][0pt]{\shortstack[l]{{\twlrm
$\varepsilon_n'$}}}}}
\put(165,160){\makebox(0,0)[lb]{\raisebox{0pt}[0pt][0pt]{\shortstack[l]{{\twlrm
$\varepsilon_2'$}}}}}
\put(170,120){\makebox(0,0)[lb]{\raisebox{0pt}[0pt][0pt]{\shortstack[l]{{\twlrm
$\varepsilon_1'$}}}}}
\put(70,390){\makebox(0,0)[lb]{\raisebox{0pt}[0pt][0pt]{\shortstack[l]{{\twlrm
$(-1)^{n+1-i}$}}}}}
\put(95,65){\makebox(0,0)[lb]{\raisebox{0pt}[0pt][0pt]{\shortstack[l]{{\twlrm
$\varepsilon$}}}}}
\end{picture}

Consider now the operator whose $(\sigma,\sigma')$ matrix element
is given by Figure 3.2.
We expect that in the large lattice limit $N\rightarrow\infty$
it will give rise to a well-defined operator
\eqa
&&\Phi_{i\,\vep}^{i+1}(\zt)~:~\H_i \goto{} \H_{i+1}. \label{eqn:VO}
\endeqa
The effect of operating with $\Phi_{i\,\vep}^{i+1}(\zt)$ amounts to inserting
a vertical dislocation in the lattice.
In view of the expression \refeq{CTM} of the CTMs
the function $F^{(i)}_n$ of \sec(1) is expressed as
\eqa
&&F^{(i)}_n(\zt_1,\cdots,\zt_n)_{\vep_1,\cdots,\vep_n}
=
{\tr_{\H_i}\left(x^{2D}\Phi^i_{i+1\,\vep_1}(\zt_1)\cdots
\Phi^{i+n-1}_{i+n\,\vep_n}(\zt_n)\right)
\over
\tr_{\H_i}\left(x^{2D}\right)
}.\label{eqn:trace}
\endeqa
%Here the suffix/superfix $i$ of
%$\Phi^{i+1}_{i\,\vep}(\zt)$ is to be read modulo $2$.

The arguments in \sec(1) can be translated to the
properties of $\Phi^{i+1}_{i\,\vep}(\zt)$.  We summarize them below.

\eqa
&&x^{2D}\circ \Phi_{i\,\vep}^{i+1}(\zt)\circ x^{-2D}
=\Phi_{i\,\vep}^{i+1}(x^2\zt), \label{eqn:deg} \\
&&\Phi^{i }_{i+1\,\vep_2}(\zt_2)
\Phi^{i+1 }_{i\,\vep_1}(\zt_1)
=\sum_{\vep'_1,\vep'_2}
R(\zt_1/\zt_2)_{\vep_1\vep_2;\vep'_1\vep'_2}
\Phi^{i}_{i+1\,\vep'_1}(\zt_1)
\Phi^{i+1}_{i\,\vep'_2}(\zt_2), \label{eqn:com} \\
&&\hbox{There exists a linear isomorphism }
\nu :\H_0 \longrightarrow \H_1 \hbox{ such that } \nnb \\
&&\qquad \qquad \nu\circ D=D\circ \nu,
\quad  \Phi_{0\,\vep}^1(\zeta) = \nu\circ \Phi_{1\,-\vep}^0(\zeta)\circ\nu, \\
&&\Phi_{i\,\vep}^{i+1}(-\zt)
=(-1)^{i+1} \vep\,\Phi_{i\,\vep}^{i+1}(\zt),
\label{eqn:par} \\
&&\sum_{\vep=\pm} \Phi_{{i+1} \,\vep}^{*i} (\zt)
\Phi_{i \,\vep}^{{i+1}} (\zt)
= \id, \quad
\Phi_{i\,\vep }^{{i+1}} (\zt)
\Phi_{{i+1}\,\vep'}^{*i} (\zt)
= \delta_{\vep\,\vep'}\times \id. \label{eqn:dual}
\endeqa
In eq.\refeq{dual} we have set
$\Phi_{i\, \vep}^{*{i+1}} (\zt)
=\Phi_{i\,-\vep}^{{i+1}}  (x^{-1}\zt)$.
The difference equations \refeq{Diff} for \refeq{trace} is an immediate
consequence of \refeq{deg}, \refeq{com}.
In the trigonometric case \refeq{VO} is related to the vertex operators
in \cite{DFJMN,JMMN} via $\Phi_{i\,\vep}^{i+1}(\zt)=g^{-1/2}\zt^{(1+\vep)/2-i}
\Phit_{\La_i\,\vep}^{\La_{i+1}}(\zt^2)$ where $g$ is a scalar independent of
$\zt$.
%It is an interesting mathematical
%problem to give an explicit realization of the
%spaces $\H_i$ and the operators \refeq{VO} with the properties
%\refeq{char,deg,com,par,dual}.

\par
\bigskip\noindent
{\it Acknowledgement.}\quad
Part of this work was begun while the authors were visiting the
Isaac Newton Institute, Cambridge, September--October 1992.
The authors would like to thank M. Atiyah, P. Goddard, the organizers
and the secretaries for their hospitality, and
B. Davies,
E. Corrigan,
V. Fateev,
O. Foda,
K. Miki,
K. Mimachi,
S. Pakuliak,
F. Smirnov
and V. Tarasov for helpful discussions.
This work is partially supported by Grant-in-Aid for Scientific Research
on Priority Areas, the Ministry of Education, Science and Culture, Japan.
\def\IJMPA{Int. J. Mod. Phys. A}
\def\CMP{Commun. Math. Phys.}
\def\LMP{Lett. Math. Phys.}
\def\NPB{Nucl. Phys. B}
\def\JSP{J. Stat. Phys.}
\def\JPC{J. Phys. C}
\def\PL{Phys. Lett.}
\def\RIMS{RIMS preprint }

\end{document}